\documentclass[aps,prb,article,twocolumn,groupedaddress,longbibliography]{revtex4-2}
\usepackage[T1]{fontenc}
\usepackage[utf8]{inputenc}
\usepackage{amsmath,amssymb,bm,mathtools}
\usepackage{graphicx}
\usepackage{xcolor}
\usepackage{hyperref}
\usepackage{physics}
\usepackage{booktabs}
\usepackage{ifthen}
\usepackage{soul}
\usepackage{cancel}
\usepackage[normalem]{ulem}
\usepackage{braket}

\hypersetup{colorlinks=true,citecolor=blue,linkcolor=blue,urlcolor=blue}

\def\rb{{\bm r}}
\def\ep{\varepsilon}
\def\ka{\varkappa}
\def\1{^{(1)}}
\def\0{^{(0)}}
\def\2{^{(2)}}
\def\3{^{(3)}}
\def\4{^{(4)}}
\def\supera{^{(\alpha)}}

\newcommand{\nh}{n_{\mathrm{h}}}
\newcommand{\todofigure}[2]{%
  \fbox{%
    \begin{minipage}[c][#2][c]{0.95\columnwidth}
      \centering
      \vspace{2mm}
      {\bfseries Figure placeholder}\\[1mm]
      #1\\[1mm]
      Replace with final vector plot file.
      \vspace{2mm}
    \end{minipage}}
}
\newcommand{\insertfig}[3]{%
  \IfFileExists{#1}{\includegraphics[width=#2]{#1}}{\todofigure{#3}{0.19\textheight}}%
}

\begin{document}

\title{Dispersive mode coupling in $p$-doped semiconductor nanomechanical resonators}

\author{Ankang Liu}
\thanks{These authors contributed equally.}
\author{Mahmoud T. Elewa}
\thanks{These authors contributed equally.}
\author{M. I. Dykman}
\affiliation{Department of Physics and Astronomy, Michigan State University, East Lansing, Michigan 48824, USA}

\date{\today}

\begin{abstract}

    Dispersive coupling between vibrational modes with different frequencies is a major nonlinear dynamical effect. We show, analytically and numerically, that in $p$-doped semiconductors such coupling can be strongly enhanced. Moreover, the relative coupling strength increases with the order of the nonlinearity. The doping-induced dispersive coupling becomes much stronger than the intrinsic one already for moderately strong doping. Its dependence on the hole density is nonmonotonic, and the temperature dependence becomes nonmonotonic for higher densities. Relevant mesoscopic frequency fluctuations are briefly discussed. The theory is applied to Si resonators, where doping is often used to compensate the temperature dependence of a clock mode, whereas another low-frequency mode is used as a thermometer to enable temperature stabilization.

\end{abstract}

\maketitle

\section{Introduction}
Micro- and nanoelectromechanical systems (MEMS and NEMS) are broadly used for precision timing, inertial sensing, and frequency control. Semiconductor-based mechanical resonators are especially attractive, because they are small, have a high quality factor $Q$, and are compatible with standard micro-scale electronics. These systems have been extensively studied both in physics and in engineering ~\cite{Cleland1996,vanBeek2012,Brand2015,Yamaguchi2017,Miller2018,Bachtold2022}. They are mesoscopic: their low-frequency eigenmodes are well-separated in frequency and can be individually accessed. At the same time, the dynamical properties are well described by macroscopic characteristics such as density and strain and stress tensors, for example.

For frequency-reference applications, in particular those that require long-term stability, the clock-mode frequency of a resonator must remain stable against comparatively slow fluctuations in the environment, including temperature and pressure fluctuations. Another factor to keep controlled is the vibration amplitude: because of the inherent nonlinearity of mechanical resonators, an amplitude change leads to a change of the vibration frequency. Whereas pressure control can be accomplished by sealing the resonator, control of temperature is often beyond direct control. Even if the eigenfrequency of a chosen vibrational mode weakly depends on temperature, it may be insufficient for the required stability. 

An efficient approach to frequency stabilization is based on a dual-mode architecture. Here, two mechanical modes of the same resonator are operated simultaneously. One mode serves as the frequency reference, while the other serves as an in situ thermometer~\cite{Ng2013,ComenenciaOrtiz2020,Kwon2020,jia2022,xiao2023,Yang2025,Yan2026}. The clock mode is chosen for its weak temperature dependence, whereas the auxiliary mode is chosen to have a larger temperature coefficient. Monitoring the auxiliary-mode frequency then provides a direct measure of the resonator temperature, which can be used to control it even where the temperature of the environment changes.

An important factor to be taken into account in the analysis of multi-mode mesoscopic vibrational system is the dispersive, or cross-Kerr, nonlinear mode-mode coupling, i.e., a nonresonant coupling between modes with incommensurate frequencies. Such coupling leads to the dependence of the mode frequency on the amplitudes of the modes it is coupled to. In particular, the clock-mode frequency depends on the amplitude of the auxiliary mode, but also on the amplitudes of other modes. To the lowest order, the corresponding coupling comes from the terms in the energy of nonresonant modes, which are bilinear in the squared modes coordinates $Q_\alpha$, where $\alpha$ enumerates the mode. Phenomenologically, the corresponding energy is $\propto Q_\alpha^2 Q_{\alpha'}^2$ with $\alpha\neq \alpha'$. 

Dispersive coupling has been well known in the physics of solids, cf.~\cite{Born1939,Stern1958}. More recently it has attracted attention in the physics of nanomechanical systems, cf.~\cite{Mahboob2015,Asadi2018,Arrangoiz-Arriola2019,Shevyrin2024,Allemeier2025,Wattjes2026}; in particular, it has been used as a sensitive probe of modal occupation~\cite{Venstra2012,Matheny2013,Gajo2020}. It plays an important role in optomechanics \cite{Aspelmeyer2014a}. Dispersive coupling of a nanomechanical modes and a qubit provides a means to measure quantum states of the mode \cite{Lee2023}. In a slightly different form, dispersive coupling underlies a commonly used way of measuring the states of superconducting qubits \cite{Blais2021}. Generally, dispersive coupling is a major mechanism of frequency fluctuations of various mesoscopic systems, including fluctuations that come from thermal fluctuations of the amplitudes of the modes dispersively coupled to the system of interest \cite{Bachtold2022}.

For all relevant applications, it is necessary to understand the strength of the dispersive coupling. In two-mode based clocks, in particular, it puts a constraint on how strongly the auxiliary mode can be driven before the frequency of the clock mode significantly changes, and thus on sensing capability of the auxiliary mode.

Dynamical properties of semiconductor-based resonators can be efficiently modified using doping and thus changing the free carrier density. This is a consequence of the typically strong electron-phonon coupling. The relevant aspect of the coupling is that the vibration-induced strain changes the electronic band structure, leading to a redistribution of the carriers within and between energy bands. The redistribution adds strain-induced terms to the carrier free energy and thus to the free energy of the crystal as whole. In turn, this changes the elasticity parameters, as they are determined by the derivatives of the free energy over strain. This is a mechanism of the electrons-mediated backaction on the vibrations. This mechanism was first proposed by Keyes~\cite{Keyes1961} and has since been analyzed in several types of semiconductors~\cite{Keyes1968,Bir1963,Hall1967,Cerdeira1972,Kim1976,Averkiev1984,Khan1985,Averkiev1987,Moskovtsev2017,Liu2025}. In silicon resonators, the carrier-mediated elastic response has been shown experimentally to significantly modify the temperature dependence of the elastic constants, reducing the temperature dependence of the frequencies of certain modes~\cite{Samarao2012,Hajjam2012} or even making it nonmonotonic~\cite{Ng2015}.

In this paper, we study the dispersive (cross-Kerr) nonlinearity in a doped semiconductor resonator. We develop a general approach, that allows us to find the strength of the coupling between low-frequency modes as a function of hole density and temperature. We also study the effect of the coupling to high-frequency modes and the frequency fluctuations that come from this coupling. The results refer to $p$-doped diamond or zinc-blende semiconductors such as Si or GaAs, which are typically used as mechanical resonators \footnote{For $p$-doped piezoelectric semiconductors with hole densities $\nh\gtrsim10^{17}$~cm$^{-3}$, the holes almost completely screen piezoelectrically induced polarization charges from low frequency modes. The piezoelectric effect then reduces to an effective electric potential for the holes, which is linear in strain. For GaAs, the corresponding hole potential energy is much smaller than the linear in the strain term in the hole chemical potential calculated in the deformation potential approximation.}. 
The branches of light and heavy holes of such semiconductors are degenerate at the center of the Brillouin zone. This makes them sensitive to strain, which lifts the degeneracy and thus significantly distorts the band structure. We extend to multi-mode nonlinear coupling the analysis of the effect of electron-phonon coupling on the self-nonlinearity of a mode \cite{Liu2025}, which was done in a single-mode approximation. This extension includes, in particular, determining the appropriate components of the fourth-order nonlinear elasticity tensor and calculating their dependence on the hole density and temperature for both degenerate and nondegenerate hole gas. 

The explicit results refer to the system of coupled modes in silicon resonators. As an example, we consider the modes studied in the experiments~\cite{ComenenciaOrtiz2020,Kwon2020,Yang2025,Yan2026}. In these and in a number of other experiments \cite{majjad2001,khine2009,holmgren2009,zhu2014}, the resonator is a thin plate, and the clock mode is a Lam\'e mode. This is a shear mode, and therefore it has reduced sensitivity to thermal expansion. It also displays a high quality factor and a low phase noise~\cite{Lee2016,Daruwalla2020}. As an auxiliary temperature-sensing mode in \cite{ComenenciaOrtiz2020}, an out-of-plane torsional mode was used. This allowed achieving parts-per-trillion-level frequency stability of the clock mode \cite{Yang2025,Yan2026}. 

In Sec.~\ref{sec:backaction} we formulate a general approach to the problem by introducing normal modes and showing how the backaction of the phonon-hole coupling modifies the elastic part of the free energy of the crystal. We also provide Table~\ref{tab:symbols} that shows major symbols used in the paper. In Sec.~\ref{sec:estimates} we provide estimates of the hole-induced parameters of linear and nonlinear elasticity. Section~\ref{sec:T_dependence_compared} provides a brief discussion of the effect of dispersive coupling of mesoscopic low-frequency modes to high-frequency phonons. Section \ref{sec:phase_diffusion} contains a discussion of phase diffusion of low-frequency modes due to dispersive mode-mode coupling. Section \ref{sec:two_modes} describes the Lam\'e and torsional modes and the expressions for the effect of their dispersive coupling on the amplitude-dependent frequency shifts in terms of the nonlinearity parameters. Section \ref{sec:three_bands} describes the numerical procedure of calculating the hole-induced contribution to the nonlinear elasticity tensor for a three-band model of the hole dispersion. Section \ref{sec:results} is the central part of the paper. It provides numerical results on the temperature and density dependence of the frequency shifts of dispersively coupled Lam\'e and torsional modes in Si. Section \ref{sec:conclusions} summarizes the results.

\begin{table*}[htbp]
\caption{\label{tab:symbols} List of Symbols}
\begin{ruledtabular}
\begin{tabular}{l l}
$\bm{u}^{(\alpha)}(\rb)$ & Dimensionless spatial displacement profile of mode $\alpha$ \\

$\ep^{(\alpha)}(\rb)$ & Dimensional spatial strain profile of mode $\alpha$ determined by $\bm{u}^{(\alpha)}(\rb)$ \\

$Q_\alpha$ & Dynamical coordinate of mode $\alpha$ \\

$\omega_{\alpha}$ & Eigenfrequency of mode $\alpha$ \\

$\hat{C}^{(n)}$ & $n$th-order intrinsic elasticity tensor \\

$\hat{\Lambda}^{(n)}$ & Doping-induced corrections to the $n$th-order elasticity tensor \\

$\eta_{\alpha}=A_{\alpha}/L_{s}$ & Amplitude $A_\alpha$ of mode $\alpha$ scaled by the side length of the plate $L_s$ \\

$\mathcal \aleph_{\alpha\beta}=\delta\omega_{\alpha}/\omega_{\alpha}\eta_{\beta}^{2}$ & Scaled frequency shift $\delta\omega_\alpha$ of mode $\alpha$ due to dispersive coupling to mode $\beta$ \\
\end{tabular}
\end{ruledtabular}
\end{table*}


\section{Adiabatic backaction from the hole-vibrational coupling}
\label{sec:backaction}

We consider a single-crystal semiconductor resonator and enumerate its vibrational modes by $\alpha$. The displacement field is 
\begin{equation}
    \label{eq:u_decomp}
    \bm{u}(\bm{r},t)=\sum_\alpha Q_\alpha(t)\bm{u}^{(\alpha)}(\bm{r}),
\end{equation}
where $Q_\alpha$ is the dynamical coordinate of the mode and $\bm{u}_\alpha$ is dimensionless mode profile normalized to the resonator volume $V$, i.e., $\int \bm{u}^{(\alpha_{1})}\cdot \bm{u}^{(\alpha_{2})} d\rb =V\delta_{\alpha_{1}\alpha_{2}}$. The corresponding strain tensor is
\begin{equation}
    \label{eq:strain_decomp}
    \hat{\varepsilon}(\bm{r},t)=\sum_{\alpha}Q_\alpha(t)  \varepsilon^{(\alpha)}(\bm{r}), \quad
    \ep^{(\alpha)}_{ij} = \left(\partial_j u^{(\alpha)}_i + \partial_i u^{(\alpha)}_j\right), 
\end{equation}
where subscripts $i,j$ enumerate the axes $x,y,z$, cf.~\cite{Landau1986}; we have disregarded quadratic in $\bm{u}^{(\alpha)}$ terms in $\ep\supera$; their contribution is much smaller than the contribution from the nonlinear terms coming from the strong electron-phonon coupling, cf.~\cite{Moskovtsev2017}. In what follows we use Voigt notation,
\begin{align*}
&\ep_1\supera =   \ep_{xx}\supera,\;\ep_2\supera =   \ep_{yy}\supera,\;\ep_3\supera =   \ep_{zz}\supera, \\
&\ep_4\supera = 2  \ep_{yz}\supera,\; \ep_5\supera = 2  \ep_{xz}\supera,\; \ep_6\supera=2  \ep_{xy}\supera.
\end{align*}

The Hamiltonian $H$ of the modes can be written as a sum of a harmonic part and nonlinear parts, $H=H_\mathrm{lin} + H_\mathrm{nl}$. For modes with inversion symmetry, $H_\mathrm{nl}$ contains only even powers of the coordinates, so that 
\begin{align}
\label{eq:Hamiltonian}
&H_\mathrm{lin} = \frac{1}{2}\sum_{\alpha}\left(\frac{P_\alpha^2}{M}+M\omega_\alpha^2 Q_\alpha^2\right) \nonumber\\
&H_\mathrm{nl} =\frac{1}{4!}\sum_{\alpha_1,\alpha_2,\alpha_3,\alpha_4}\gamma^{(\alpha_1\alpha_2\alpha_3\alpha_4)} Q_{\alpha_1}Q_{\alpha_2}Q_{\alpha_3}Q_{\alpha_4} + \ldots,
\end{align}
where $M$ is the mass of the resonator, $\omega_\alpha$ is the eigenfrequency of mode $\alpha$, and $\hat\gamma$ is the tensor of the 4th order nonlinearity. Here and below we imply summation over repeated indices. The dispersive coupling of modes $\alpha_1$ and $\alpha_2$ is described by the term $(1/4) \gamma ^{(\alpha_1\alpha_1\alpha_2\alpha_2)}Q_{\alpha_1}^2 Q_{\alpha_2}^2$ in $H_\mathrm{nl}$. The manifestation of this coupling in the mode dynamics is discussed in Sec.~\ref{sec:phase_diffusion} below.

We will be interested in the effect of doping on the modes with frequencies $\omega_\alpha$ in the range of $2\pi\times 10^6$ -- $2\pi\times 10^9$~s$^{-1}$. This range covers the typical frequencies of nano- and micro-mechanical resonators.

The analysis below directly extends to the modes with no inversion symmetry. For such modes, one has to take into account terms which are cubic in $Q_\alpha$. As we show below, the doping-induced contribution from these terms to the dispersive coupling parameters is small.

\subsection{The contribution of the hole-phonon coupling to the elastic free energy: general formulation}
\label{subsec:general_elastic}

For diamond- and zinc-blende semiconductors, the valence band is formed by $p$-orbital atomic states, and the analysis should take into account three branches of the valence band. They correspond to heavy and light holes and the band, which is split off by the spin-orbit coupling.
To describe the effect of the hole-vibrational coupling we note that, for the hole densities and mode frequencies of interest, the hole thermalization time is much shorter than the reciprocal frequencies. The hole subsystem can be then described in the adiabatic approximation, as it remains in local quasiequilibrium for a given instantaneous strain field~\cite{Liu2025}. The density of the holes $\nh$ is determined by the acceptor density. If this density is high and the temperature is not too low, all acceptors are ionized and the hole density is equal to the density of the acceptors. For Si, for example, this happens in the range where the acceptor densities are above $10^{18}~\text{cm}^{-3}$ and $T\gtrsim 100$~K. 

In the adiabatic approximation, the hole grand canonical potential density is 
\begin{equation}
\Omega = -2k_B T\sum_{\nu}\int \frac{d^3\bm{k}}{(2\pi)^3} \ln\! \left[1+\exp\!\left(\frac{\mu-E_\nu(\bm{k},\hat{\varepsilon})}{k_B T} \right)\right],
\label{eq:grand_potential}
\end{equation}
where the factor of two accounts for the Kramers degeneracy, $\nu$ labels the energy branches of the valence band, and $E_\nu(\bm{k},\hat{\varepsilon})$ is the hole energy. It can be found for a given strain $\hat\ep$ by diagonalizing the Luttinger--Kohn--Bir--Pikus Hamiltonian~\cite{Luttinger1955,Bir1974,Winkler2003}. The hole free energy density is $\mathcal{F}_{\text{h}} = \Omega +\mu \nh$, where $\mu\equiv \mu(\hat\ep)$ is the chemical potential.

The strain-induced term in the hole free-energy density $\Delta\mathcal{F}$ can be expanded in a power series in the strain,
\begin{align}
    \label{eq:free_energy_expand}
    \Delta\mathcal{F} =& \frac{1}{2}\Lambda^{(2)}_{ij}\varepsilon_{i}\varepsilon_{j} +\frac{1}{3!}\Lambda^{(3)}_{ijk}\varepsilon_{i}\varepsilon_{j}\varepsilon_{k}\nonumber\\
    &+\frac{1}{4!}\Lambda^{(4)}_{ijkl}\varepsilon_{i}\varepsilon_{j}\varepsilon_{k}\varepsilon_{l}+\cdots,
\end{align}
where $\ep_i$ are the components of the strain tensor $\hat\ep$ in the Voigt notations.

The term $\propto\Lambda\2$ in Eq.~(\ref{eq:free_energy_expand}) has the same form as the elastic part of the free energy of the host crystal in the harmonic approximation. It directly gives the changes of the modes' eigenfrequencies. These changes depend on temperature, as seen from Eq.~\eqref{eq:grand_potential}. The term $\propto\Lambda\4$ taken to the first order determines the parameters $\gamma^{(\alpha_{1}...\alpha_{4})}$ of the quartic nonlinearity in $H_\mathrm{nl}$. 


\section{Estimates of the coupling-induced elasticity parameters}
\label{sec:estimates}

Prior to analyzing the effects of dispersive coupling on the dynamics of low-frequency modes and comparing the effects of doping and the intrinsic nonlinearity of the host crystal, we need to estimate the magnitude of the doping-induced terms in the elastic free energy. It follows from Eqs.~\eqref{eq:grand_potential} and \eqref{eq:free_energy_expand} that the $n$th-order hole-induced elastic coefficients scale as
\begin{equation}
\label{eq:Lambda_n_est}
\|\hat\Lambda^{(n)}\| \sim \nh \frac{\mathcal{D}^{n}}{E_{\mathrm{kin}}^{\,n-1}},
\qquad n\ge 2,
\end{equation}
cf. Ref.~\cite{Liu2025}. Here $\mathcal{D}$ is the typical value of the deformation potential parameter of the material and
\begin{equation}
\label{eq:E_kin}
E_{\mathrm{kin}}= \max(k_B T,\mu)
\end{equation}
is the characteristic carrier kinetic energy, which crosses over from $k_B T$ in the nondegenerate regime to $\mu$ in the strongly degenerate regime. For semiconductors, $\mathcal{D}$ is typically of the order of a few electron-volts, whereas $E_{\mathrm{kin}}$ is $\sim 10^{-2}$--$10^{-1}$~eV in the relevant range of temperatures and hole densities. Thus $\mathcal{D}/E_{\mathrm{kin}}\gg 1$. 

Even though $\mathcal{D}/E_{\mathrm{kin}}$ is large, the quadratic in $\hat\ep$ term in $\Delta\mathcal{F}$ is small compared with the quadratic in $\hat\ep$ term in the free energy of the host crystal. The latter term has the form $\frac{1}{2} C\2_{ij}\varepsilon_i \varepsilon_j$. The parameters of the ``acoustic part'' of the intrinsic linear elasticity tensor $\hat C\2$, including the Young modulus, are $\|\hat C\2\|\sim \rho v_s^2$, where $\rho$ is the crystal density and $v_s$ is the sound velocity. The parameter ratio that characterizes the relative magnitude of the quadratic in $\hat\ep$ term in $\Delta\mathcal{F}$ is
\begin{equation}
\label{eq:smallness}
\frac{\|\hat\Lambda^{(2)}\|}{\|\hat C\2\|}
\sim
\frac{\nh \mathcal{D}^{2}}{E_{\mathrm{kin}}\rho v_s^{2}}.
\end{equation}
For typical values $\mathcal{D}\sim 3~\mathrm{eV}$, $\nh\sim 10^{19}~\mathrm{cm}^{-3}$, $E_{\mathrm{kin}}\sim 0.03~\mathrm{eV}$, $\rho\sim 2.3~\mathrm{g\,cm^{-3}}$, and $v_s\sim 5\times 10^{5}~\mathrm{cm\,s^{-1}}$, it is $\sim 10^{-2}$. 

An extra small factor that affects the ratio of the hole-induced elasticity to the intrinsic one comes from the fact that the adiabatic description of the hole-vibrational coupling used in Eq.~\eqref{eq:free_energy_expand} applies to vibrations with frequencies small compared to the reciprocal hole thermalization time. If the thermalization time is $t_{\text{h}}\sim 10^{-12}$~s, the wave vectors of the relevant acoustic modes are $\lesssim q_{\max} \sim (v_s t_{\text{h}})^{-1} \sim 2\times 10^6$~cm$^{-1}$. This value is small compared to the wave vector of thermal phonons $q_T\sim k_BT/\hbar v_s\gtrsim 3\times 10^7$~cm$^{-1}$ for $T>100$~K.

In contrast to the quadratic in $\hat\ep$ part of $\Delta\mathcal{F}$, because of the large value of $\mathcal{D}/E_{\mathrm{kin}}$, the quartic in $\hat\ep$ terms are typically large compared to the corresponding intrinsic terms for the modes with frequencies smaller than $1/t_{\text{h}}$. This makes it important to calculate them and to find their dependence on the hole density and temperature.

\subsection{Convergence of the free energy expansion in strain}

The expansion (\ref{eq:free_energy_expand}) is justified if the terms of higher order in $\hat \ep$ are smaller than the lower-order terms.
It is convenient to compare the quartic and quadratic terms in Eq.~\eqref{eq:free_energy_expand} by comparing their thermal averages integrated over the volume. Decoupling $\braket{\hat\ep\otimes\hat\ep\otimes\hat\ep\otimes\hat\ep} \sim \braket{\hat\ep\otimes\hat\ep}\braket{\hat\ep\otimes\hat\ep}$, we obtain, to the order of magnitude, 
\begin{align}
\label{eq:ratio}
&R=\frac{\int d\rb \hat\Lambda\4 \braket{\hat\ep\otimes\hat\ep\otimes\hat\ep\otimes\hat\ep}}{\int d\rb\hat\Lambda\2\braket{\hat\ep\otimes\hat\ep}} \nonumber\\ 
&\sim V^{-1}
\frac{\|\hat\Lambda^{(4)}\|}{\|\hat\Lambda^{(2)}\|^2 }\int d\rb\hat\Lambda\2\braket{\hat\ep\otimes\hat\ep} \nonumber\\
&\sim (V \nh E_\mathrm{kin})^{-1} \int d\rb\hat\Lambda\2\braket{\hat\ep\otimes\hat\ep},
\end{align}
where $V$ is the volume of the system. Thermal average of the squared strain tensor can be estimated using the standard expansion for $\hat\ep$ in terms of normal vibrational modes, cf.~\cite{Girvin2019}. For classical vibrations, $k_BT\gg \hbar v_s q_{\max}$, this gives 
\[\int d\rb\braket{\hat\ep\otimes\hat\ep} \sim \frac{k_BT V}{\pi^2\rho v_s^2}q_{\max}^3.
\]
In turn, for the ratio \eqref{eq:ratio}, this gives 
\begin{align}
    \label{eq:R_estimate}
R\sim \frac{\|\hat\Lambda\2\| k_BT}{\pi^2\rho v_s^2 E_\mathrm{kin} \nh}q_{\max}^3 \lesssim 10^{-3}.
\end{align}
Similar estimates can be made for higher-order terms in the expansion \eqref{eq:free_energy_expand} of the adiabatic part of the hole-induced terms in the elastic free energy. They justify the perturbation theory in $\hat\ep$ used to obtain Eq.~\eqref{eq:free_energy_expand}.

\section{The effect of dispersive coupling to high-frequency phonons}
\label{sec:T_dependence_compared}

We now estimate the contributions to the temperature dependence of the eigenfrequencies of low-frequency eigenmodes of a mechanical resonator. They come from both intrinsic elastic properties of the crystal and the effect of $p$-doping. They also come from phonons with different frequencies. In this section we will separate and estimate relative magnitude of different contributions. The starting point is that, in the harmonic approximation, the squared eigenfrequency of the mode $\alpha$ is determined by the sum of the linear elasticity components with and without doping, 
\begin{align}
\label{eq:lin_elasticity}
    \omega_{\alpha}^{2}=M^{-1}\int d\rb \left(C_{ij}^{(2)}+\Lambda_{ij}^{(2)}\right)\varepsilon_{i}^{(\alpha)}\varepsilon_{j}^{(\alpha)}.
\end{align}
In the absence of doping, the temperature dependence of eigenfrequencies comes from the intrinsic vibration nonlinearity. The free energy density of the vibrations of the host crystal has the form of the series \eqref{eq:free_energy_expand}, with $\hat\Lambda^{(n)}$ replaced by the intrinsic elasticity tensors $\hat C^{(n)}$ with $n=2,3,4,...$. The tensors $\hat C^{(n)}$ come from the expansion of the potential energy of the interatomic coupling in a crystal. Generally, their major components are of the same order of magnitude, at least for not too large $n$ \cite{Gurevich1988} (however, a numerical approach \cite{Pandit2023} predicted $\hat C^{(n)}$ with the components that increase with $n$, with $\|\hat{C}^{(4)}\|/\|\hat{C}^{(2)}\|\gtrsim 10$ for Si, in particular; the physical origin of this increase was not specified). For our estimate we will use $\|\hat C\2\| \sim \|\hat C\3\|\sim \|\hat C\4\|$. 

To the second order in $\hat C\3$ and to the first order in $\hat C\4$, the linear elasticity parameters of $\hat C\2$ in Eq.~(\ref{eq:lin_elasticity}) acquire temperature-dependent corrections $\sim 
(\|\hat C\3\|^2/\|\hat C\2\|)\braket{\hat\ep\otimes\hat\ep},
\|\hat C\4\|\braket{\hat\ep\otimes\hat\ep}$, cf. \cite{Born1939,Stern1958,Garber1975,Malica2020} (see also Sec.~\ref{sec:phase_diffusion} below). For room temperatures, these corrections are determined primarily by short-wavelength thermal phonons, as these phonons have a large density of states. The relative magnitude of the corrections is 
\begin{align}
\label{eq:intrinsic_T_dependence}
\left(\frac{\|\hat{C}^{(4)}\|\braket{\hat\ep\otimes\hat\ep}}{\|\hat{C}^{(2)}\|}\right)_\mathrm{intrinsic} \sim \frac{k_{B}T q_{T}^{3}}{\pi^2\rho v_{s}^{2}} \lesssim 10^{-2}.
\end{align}
Here $q_T=k_B T/\hbar v_s$ is the thermal wave number. This estimate gives $|d\ln\omega_\alpha^2/d\ln T|_\mathrm{intrinsic} \sim 10^{-2}$. For silicon, in particular, such estimate agrees with the results of various studies, cf.~\cite{Mcskimin1953,Hall1967,Varshni1970,Burenkov1974}. 

Because of doping, the frequency becomes temperature-dependent even in the harmonic approximation \eqref{eq:lin_elasticity}. It follows from Eqs.~\eqref{eq:Lambda_n_est} -- \eqref{eq:smallness} that the temperature dependence of the doping-induced correction to the frequency in the linear-elasticity approximation is 
\begin{align}
    \label{eq:T_estimate}
\left|\frac{d\ln \omega_\alpha^2}{d\ln T}\right|_\mathrm{doping} 
\sim \|\hat C\2\|^{-1}\frac{d\|\hat\Lambda\2\|}{d\ln T} \sim 10^{-2}\text{--} 10^{-3}.
\end{align}
It is comparable to the intrinsic temperature dependence of the frequency, which explains the success of using doping to compensate the intrinsic effect.

Even though $\|\hat \Lambda\2\| \ll \|\hat C\2 \|$, the estimate of the hole-induced nonlinearity \eqref{eq:Lambda_n_est} gives
\[
\frac{\|\hat \Lambda\4\|}{\|\hat C\4\|} \sim \frac{n_h\mathcal{D}^4/E_\mathrm{kin}^3}{\rho v_s^2 } \gtrsim 10^2.
\]
This shows that the doping-induced nonlinearity is much stronger than the intrinsic nonlinearity of a crystalline resonator. 

However, the density of states of thermal phonons admixed to low-frequency modes by the intrinsic nonlinearity is much larger than the density of states of phonons admixed by the adiabatic hole-phonon coupling. The thermal wave number $q_T=k_B T/\hbar v_s$ in the expression \eqref{eq:intrinsic_T_dependence}
for the intrinsic temperature-dependent frequency correction is much larger than the wave number $q_{\max}\sim (v_st_\mathrm{h})^{-1}$. As a result, even though $\|\hat \Lambda\4\| \gg\|\hat C\4\|$ for the coupling to phonons with $q\lesssim q_\mathrm{max}$, the temperature-dependent correction to $\omega_\alpha$ due to the doping-induced nonlinearity is much smaller than that due to the intrinsic nonlinearity. It is also much smaller than the correction to the temperature dependence from the doping-induced terms in the linear elasticity, cf. Sec.~\ref{sec:results}. We note that $T$-dependent frequency shift comes in the second order of the perturbation theory from the cubic in $\ep$ terms in the free energy. Again, for $q_T\gg q_\mathrm{max}$, the main contribution to the shift comes from the intrinsic nonlinearity of the crystal.


\section{Phase diffusion due to dispersive coupling}
\label{sec:phase_diffusion}

Since the doping-induced nonlinear coupling between low-frequency modes is much stronger than the intrinsic coupling, to describe this coupling we can approximate the quartic nonlinearity parameters in Eq.~(\ref{eq:Hamiltonian}) by keeping only the doping-induced term. The quartic-nonlinearity term has the form
\begin{align}
\label{eq:gamma_via_Lambda}
\gamma^{(\alpha_{1}\alpha_{2}\alpha_{3}\alpha_{4})}=\int d\rb\,\Lambda_{ijkl}^{(4)}\ep_{i}^{(\alpha_{1})}\ep_{j}^{(\alpha_{2})}\ep_{k}^{(\alpha_{3})}\ep_{l}^{(\alpha_{4})}.
\end{align}

The term in the nonlinear Hamiltonian $H_\mathrm{nl}$, Eq.~\eqref{eq:Hamiltonian}, that describes dispersive coupling of modes $\alpha$ and $\beta$ is bi-quadratic in the mode coordinates, 
\[H_\mathrm{nl}^{(\alpha\beta)}=\frac{1}{4}\gamma^{(\alpha\alpha\beta\beta)}Q_\alpha^{2}Q_\beta^{2}.\]
This term leads to a frequency shift of mode $\alpha$ that is proportional to the squared amplitude of mode $\beta$, and vice versa. Assuming that each of these modes is sinusoidally modulated in time, $Q_{\alpha}(t)=A_{\alpha}\cos(\omega_{\alpha}t+\phi_{\alpha})$, we find from the equations of motion $\ddot Q_\alpha = -\omega_\alpha^2 Q_\alpha - M^{-1}\partial_{Q_\alpha}H_\mathrm{nl}$ that the dispersive frequency shifts are, respectively
\begin{align}
\label{eq:freq_shift}
\delta\omega_{\alpha}=\frac{\gamma^{(\alpha\alpha\beta\beta)}}{8M\omega_{\alpha}}A_{\beta}^{2}, 
\quad \beta\neq \alpha.
\end{align}

Equation~\eqref{eq:freq_shift} immediately extends to the coupling of a low-frequency mode to thermal phonons,
\begin{align}
\label{eq:random_shift_general}
\delta\omega_\alpha =\frac{1}{8M\omega_\alpha}\sum_\ka \gamma^{(\alpha\alpha\ka\ka)} A_\ka^2,
\end{align}
where $\ka$ enumerates thermal phonons and $A_\ka$ is the vibration amplitude of phonon $\ka$. To describe the coupling to short-wavelength phonons, which are not adiabatically followed by the hole gas, the expression for $\hat\gamma$ in Eq.~(\ref{eq:random_shift_general}) should include contributions from both intrinsic and doping-induced quartic nonlinearity. This means that, in Eq.~(\ref{eq:gamma_via_Lambda}), one should replace $\hat\Lambda\4 \to \hat\Lambda\4 + \hat C\4$ (the parameters $\hat\Lambda\4$ and $\hat C\4$ are renormalized by the cubic nonlinearity parameters, in this case). 

Equation~\eqref{eq:random_shift_general} for the thermal-averaged frequency shift $\braket{\delta\omega_\alpha}$ describes the thermal shift of the mode frequency due to phonon nonlinearity, with $\braket{A_\ka^2}=2k_BT/M\omega_\ka^2$ for thermal vibrations. Note the scaling with the volume of the system $V$: with the chosen normalization, $\gamma^{(\alpha\alpha\ka\ka)}$ scales as $V$, whereas the squared amplitude of thermal phonons $A_\ka^2 \propto 1/M\propto 1/V$. To allow for quantum effects, one should replace $A_\ka^2 \to (2\hbar/M\omega_\ka)\hat n_\ka$, with $\hat n_\ka$ being the occupation number operator of phonon $\ka$. The $T=0$-part of the energy of nonlinearly coupled vibrations is standardly incorporated into the mode eigenfrequencies, cf.~\cite{Dykman1973a}.

The amplitudes $A_\ka$ in Eq.~\eqref{eq:random_shift_general} are random. Therefore, along with the $T$-dependent shift $\braket{\delta\omega_\alpha}$, there occur frequency fluctuations. For a mode $\alpha$ that performs self-sustained vibrations, it is convenient to describe the effect of these fluctuations by looking at the dispersive-coupling induced diffusion of the vibration phase $\phi_\alpha$. In turn, this comes to calculating the phase variance that accumulates over time $t$,
\begin{align}
    \label{eq:phase_diff_general}
\braket{\Delta\phi_\alpha^2(t)} = \braket{\left[\int_0^t dt' \bigl(\delta\omega_\alpha (t') - \braket{\delta\omega_\alpha}\bigr)
\right]^2}. 
\end{align}

Since $\delta\omega_\alpha$ is a sum over the modes $\ka$, as seen from Eq.~(\ref{eq:random_shift_general}), the phase variance $\braket{\Delta\phi_\alpha^2(t)}$ is given by a double sum over $\ka, \ka'$. The modes $\ka$ and $\ka'$ with different $\ka$ and $\ka'$ are statistically independent. Therefore, the calculation comes to evaluating the averages
\begin{align*}
    &\braket{\iint_0^t dt_1\, dt_2 [A_\ka^2(t_1)A_\ka^2(t_2) - \braket{A_\ka^2}^2]}\\ 
    &=2\int_0^t dt_1 \int_0^{t_1} dt_2 \braket{A_\ka^2(t_1)[A_\ka^2(t_2) - \braket{A_\ka^2}]}.
\end{align*}
Decay of the phonon amplitudes is typically exponential \cite{Peierls1955}, and given the Boltzmann distribution of the squared amplitudes, we obtain 
\[
\braket{A_\ka^2(t_1)[A_\ka^2(t_2) - \braket{A_\ka^2}]}
=\braket{A_\ka^2}^2\exp(-|t_1 - t_2|/\tau_\ka),
\]
where $\tau_\ka^{-1}$ is the decay rate of mode $\ka$. This gives, for $t$ much larger than the phonon relaxation times,
\begin{align}
    \label{eq:phase_variance_explicit}
 \braket{\Delta\phi_\alpha^2(t)} \approx 2D^\mathrm{ph}_\alpha t,\quad  D^\mathrm{ph}_\alpha=\sum_\ka\left(\frac{\gamma^{(\alpha\alpha\ka\ka)} k_BT}{4M^2\omega_\alpha\omega_\ka^2}\right)^2 \tau_\ka.   
\end{align}

The dispersive-coupling induced phase diffusion (\ref{eq:phase_variance_explicit}) is a mesoscopic effect. Since $M=\rho V$, whereas summation over $\ka$ gives only a factor $\propto V$, the diffusion coefficient $D^\mathrm{ph}_\alpha$ goes to zero in the thermodynamic limit $V\to \infty$. An estimate of the contribution of thermal phonons shows that it is extremely small in micromechanical systems with volumes $\gtrsim 10^{-4}~\text{cm}^3$. However, it becomes larger for nanoscale systems.

We note that, if decay of thermal phonons was disregarded and one just calculated $\braket{(\delta\omega_\alpha)^2} - \braket{\delta\omega_\alpha}^2$, one would obtain a fairly strong constraint on the frequency stability of microelectromechanical systems. 

The frequency shift $\delta\omega_\alpha$ of a low-frequency mode $\alpha$, as given by Eq.~(\ref{eq:random_shift_general}), becomes much larger if the mode $\beta$ is excited and performs forced vibrations with amplitude $A_\beta$ much larger than the typical thermal amplitude $(k_BT/M\omega_\beta^2)^{1/2}$. This is why it is interesting and important to calculate the dispersive coupling parameters.

\subsection{Renormalization of the quartic nonlinearity parameters for low-frequency modes}

\label{subsubsec:cplng_low_freq}

Generally, the parameters $\gamma^{(\alpha_1...\alpha_4)}$ are renormalized by the cubic in $\hat\ep$ terms in the free energy density. The corresponding hole-induced contribution to these terms has the form
\begin{align*}
&\Delta\mathcal{F}\3 = \frac{1}{3!}\sum \beta^{(\alpha_1\alpha_2\alpha_3)}Q_{\alpha_1}Q_{\alpha_2}Q_{\alpha_3}, \\
&\beta^{(\alpha_1\alpha_2\alpha_3)}
=\int d\rb\,\Lambda_{ijk}^{(3)}\ep_{i}^{(\alpha_{1})}\ep_{j}^{(\alpha_{2})}\ep_{k}^{(\alpha_{3})}.
\end{align*}
General expression for the renormalization of the parameters $\gamma^{(\alpha_1...\alpha_4)}$ that define the intrinsic and hole-induced mode nonlinearity by the parameters $\beta^{(\alpha_1...\alpha_3)}$ is given in Refs.~\cite{Dykman1971,Moskovtsev2017}. The change of $\|\hat\gamma\|$ is $\propto \|\hat\beta\|^2$. To obtain an estimate, we note that, for acoustic modes with typical wavelength $\lambda$, and thus with frequencies $\sim v_s/\lambda$, we have $\|\hat\gamma\| \sim \|\hat\Lambda\4\|V/\lambda^4 $ and
$\|\hat\beta\| \sim \|\hat\Lambda\3\|V/\lambda^3 $. Then the ratio of the renormalization $\|\delta\hat\gamma\|$ to $\|\hat\gamma\|$ is
\begin{align*}
\frac{\|\delta\hat\gamma\|}{\|\hat\gamma\|}
\sim \frac{\|\hat\beta\|^2}{M(v_s/\lambda)^2\|\hat\gamma\|} \sim 
\frac{\|\hat\Lambda\3\|^2}{\|\hat\Lambda\4\|\rho v_s^2}
\sim \frac{n_h \mathcal{D}^2}{E_\mathrm{kin}\rho v_s^2} \ll 1.
\end{align*}
In what follows, the renormalization $\delta\hat\gamma$ is disregarded. The above estimate refers to $\hat\gamma, \hat\beta$ for $\{\alpha_i\}$ enumerating long-wavelength acoustic modes, but it applies also to the torsional mode considered below.

We note that the effective doping-induced increase of the scaled nonlinearity parameters with the order of nonlinearity allows one to use mode-mode coupling to modulate not only the eigenfrequency of one mode by changing the amplitude of another mode, but also to modulate the mode Duffing nonlinearity parameter. The corresponding term in the Hamiltonian for modes $\alpha$ and $\beta$ is $\propto Q_{\alpha}^{4}Q_{\beta}^{2}$. This effect, along with other effects of higher-order nonlinearity, is beyond the scope of current paper.

As mentioned above, the ``intrinsic'' cubic terms of the form $Q_\alpha Q_\ka Q_{\ka'}$, with $\ka,\ka'$ referring to thermal phonons, should not be disregarded, as they lead to a temperature dependence of the eigenfrequencies of low-frequency modes $\alpha$. Formally, for a mode $\alpha$, this dependence is described by the real part of the self-energy, the imaginary part of which determines the decay rate of this mode. Generally, doping can change this self-energy, as it affects the lifetime of high-frequency phonons. However, the effect of this change on the temperature dependence of the eigenfrequencies of low-frequency modes should be small, provided these modes have small relaxation rates, which has been assumed to be the case.

\section{Dispersive coupling of two low-frequency modes}
\label{sec:two_modes}

We now apply the general results to dispersive coupling between the modes with wavelengths of the order of the crystal size. Numerical results will refer to thin square plates of lateral size $L_s$ in the range $10\text{ -- }500~\mu$m and thickness $h_s$ in the range $10\text{ -- }100~\mu$m, although the general analysis is not limited to this size range. The typical wavelengths of the modes will be $\sim L_s$. In doped micromechanical systems, such modes have frequencies much smaller than the hole relaxation/thermalization rates $1/t_{\text{h}}$, and the mechanical vibrations are adiabatically followed by the holes. 

The modes we consider are mesoscopic: they are described by continuum elasticity, but their frequencies are well-separated. It is convenient to scale their amplitudes $A_\alpha$ and introduce a dimensionless parameters $\eta_\alpha = A_\alpha/L_s$. The strength of the mode dispersive coupling is characterized by the dimensionless parameters 
\begin{align}
\label{eq:scaled_shifts_basic}
\mathcal \aleph_{\alpha\beta} =\frac{\delta\omega_\alpha}{\omega_\alpha\eta_\beta^2}
 \quad(\alpha\neq \beta), \quad \eta_\alpha = A_\alpha/L_s.
\end{align}
Equation~\eqref{eq:scaled_shifts_basic} shows that, although the dispersive interaction itself is reciprocal, $\gamma^{(\alpha\alpha\beta\beta)} = \gamma^{(\beta\beta\alpha\alpha)}$, the observable fractional frequency shifts are not equal even for equal mode amplitudes. This should not be confused with the nonreciprocity in systems away from thermal equilibrium.

Of primary interest is the dependence of the dispersive coupling parameters on the temperature and hole density for different semiconducting systems. In the remainder of the paper, we specialize the general two-mode framework to the mode pair studied in the experiment \cite{ComenenciaOrtiz2020,Yang2025,Yan2026}. The clock mode $\alpha$ is the second Lam\'e mode and the mode $\beta$ is the torsional mode of a square $p$-doped silicon plate.


\begin{figure}[t]
\centering
\includegraphics[width=5cm]{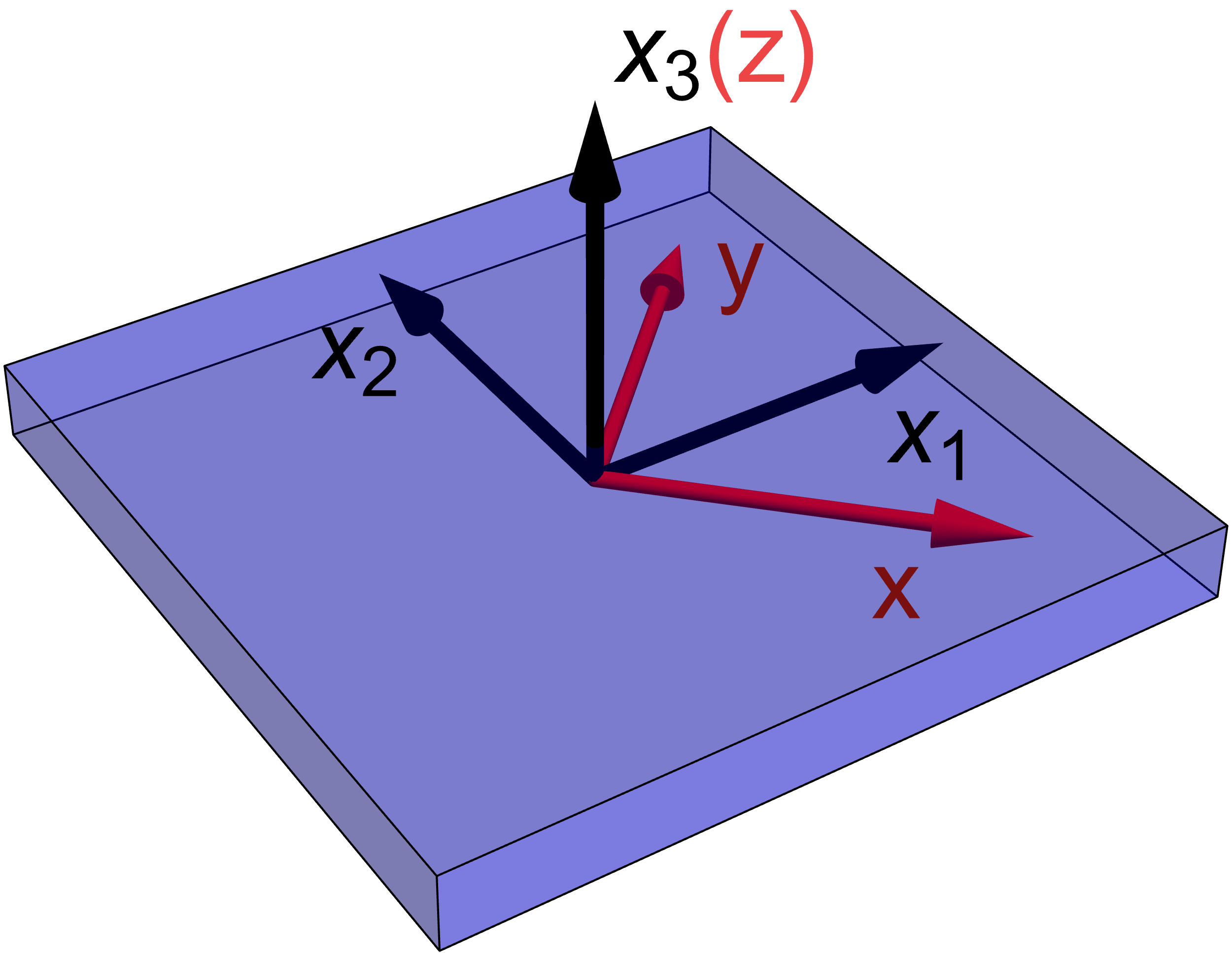}
\caption{The coordinate system for a thin square plate with side length $L_s$ and thickness $h_s$. For a plate cut out along the $\braket{100}$ crystalline axes, the coordinates $(x_{1},x_{2},x_{3})$ are parallel to the axes, as indicated by the black arrows. For a plate cut out in the plane along the $\langle110\rangle$ axes, the coordinates $(x,y)$ are shown by the red arrows, while the $z$-coordinate is still normal to the plate and points along $\braket{100}$ axes.}
\label{fig:geometry}
\end{figure}

\subsection{Lam\'e and torsional mode profiles}
\label{subsec:mode_shapes}

We consider a thin silicon square plate of side length $L_s$ and thickness $h_s \ll L_s$. We assume that the plate is cut out either along $\braket{100}$ or $\braket{110}$ axes. The coordinate system is shown in Fig.~\ref{fig:geometry}. Two coordinates [$(x,y)$ or $(x_1,x_2)$, depending on the axes orientation] lie in the midplane of the plate, whereas the third coordinate, $x_3$, is normal to the plate. The origin is chosen at the center of the plate.

The second Lam\'e mode is an isochoric in-plane mode. In the thin-plate limit, its displacement field with components $(u_1,u_2, u_3)$ is independent of $x_3$ and is given by \cite{Graff1991}
\begin{align}
    \label{eq:lame_mode}
    u_1^{(\mathrm{L})}(x_1,x_2) &= \sqrt{2}\cos\left(\frac{2\pi x_1}{L_s}\right)\sin\left(\frac{2\pi x_2}{L_s}\right), \nonumber\\
    u_2^{(\mathrm{L})}(x_1,x_2) &= -\sqrt{2}\sin\left(\frac{2\pi x_1}{L_s}\right)\cos\left(\frac{2\pi x_2}{L_s}\right)
\end{align}
with $u_3^{(\mathrm{L})}=0$. This displacement field satisfies $\partial_{x_1}u_1^{(\mathrm{L})}+\partial_{x_2}u_2^{(\mathrm{L})}=0$, so that the corresponding linear strain is traceless.

The torsional mode is described within the Kirchhoff--Love plate theory by a transverse displacement $w^{(\mathrm{T})}(x_1,x_2)$ of the midplane and the accompanying in-plane bending displacements (see also Ref.~\cite{Landau1986}),
\begin{align}
    \label{eq:torsional_mode}
    u_{1}^{(\mathrm{T})}(x_{1},x_{2},x_{3}) &= -x_{3}\,\partial_{x_{1}} w^{(\mathrm{T})}(x_{1},x_{2}),\nonumber\\
    u_{2}^{(\mathrm{T})}(x_{1},x_{2},x_{3}) &= -x_{3}\,\partial_{x_{2}} w^{(\mathrm{T})}(x_{1},x_{2}),\nonumber\\
    u_{3}^{(\mathrm{T})}(x_{1},x_{2},x_{3}) &= w^{(\mathrm{T})}(x_1,x_2).
\end{align}
For the square plate geometry considered here, the leading-order torsional mode shape may be written as \cite{Leissa1969}
\begin{equation}
    \label{eq:transverse_disp}
    w^{(\mathrm{T})}(x_{1},x_{2}) = \frac{12}{\sqrt{1+2h_s^2/L_s^2}}
    \left(\frac{x_1 x_2}{L_s^2}\right).
\end{equation}

\begin{figure}[htp!]
	\centering
	\includegraphics[width=7cm]{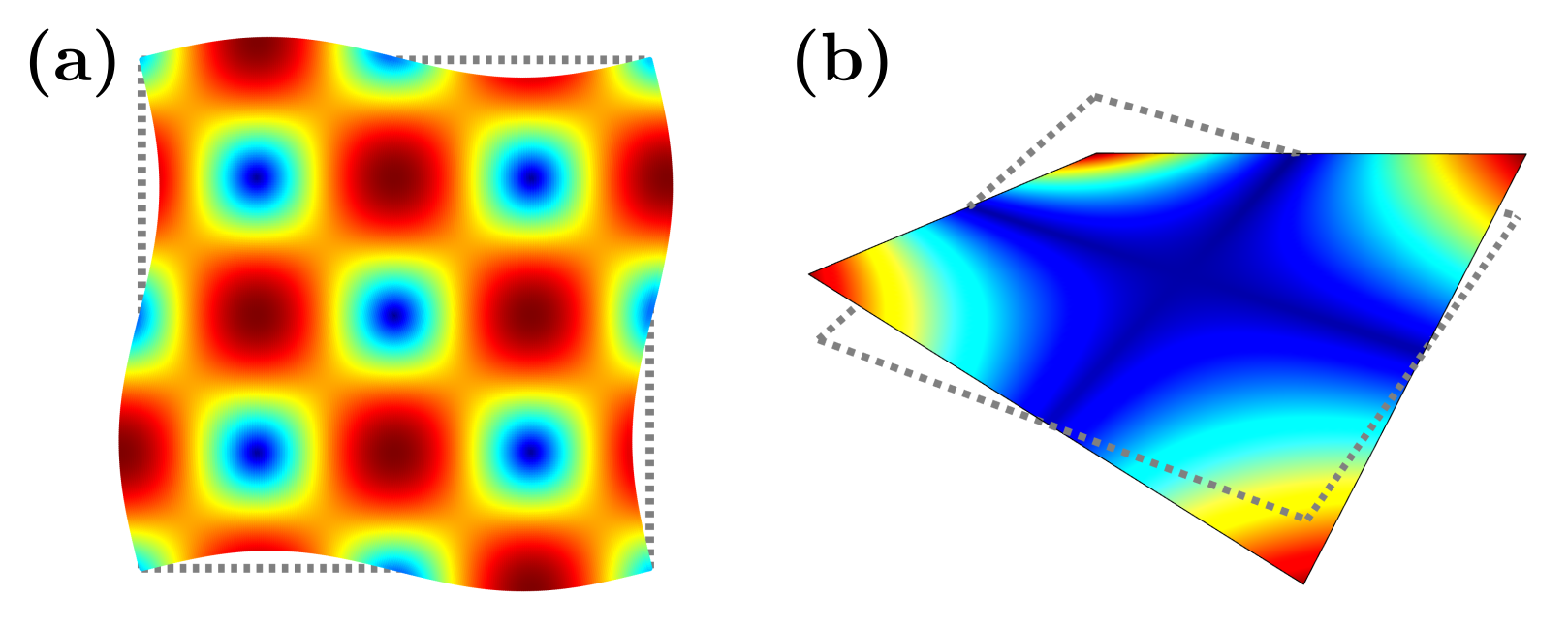}
	\caption{Mode profiles for (a) the in-plane second Lam\'e mode and (b) the considered torsional mode. The modes can be simultaneously activated in a thin square plate that is clamped through the middle points of its four sides.}
	\label{fig:mode}
\end{figure}

With the prefactors chosen in Eqs.~\eqref{eq:lame_mode} and \eqref{eq:transverse_disp}, both profiles satisfy the normalization condition
\[\frac{1}{V}\int d\rb\,|\bm{u}^{(\mathrm{L})}|^2 = \frac{1}{V}\int d\rb\,|\bm{u}^{(\mathrm{T})}|^2 = 1.\]
The displacement profiles for the modes are sketched in Fig.~\ref{fig:mode}.


\subsubsection{Mode displacement in the $\braket{100}$-basis}
\label{subsec:coordinates}

The elastic tensor components of single-crystal silicon are conventionally expressed in the coordinate system where the axes $(x,y,z)$ are pointing along the $\braket{100}$-axes. Therefore, if the square plate is cut out so that its edges are along the $\langle100\rangle$ directions, the strain tensor can be computed directly from the displacement fields in Eqs.~\eqref{eq:lame_mode} and \eqref{eq:torsional_mode}.

For a different in-plane crystal orientation, it is convenient to rotate the displacements to the $\braket{100}$ axes. If the plate edges are aligned with the in-plane $\langle110\rangle$ directions, the coordinate system $(x_1,x_2,x_3)$ has to be rotated by $\pi/4$, as shown in Fig.~\ref{fig:geometry},
\begin{equation*}
    x_{1}=\frac{x+y}{\sqrt{2}},
    \quad
    x_{2}=-\frac{x-y}{\sqrt{2}},
    \quad
    x_{3}=z.
\end{equation*}
Respectively, the displacement components become
\begin{equation}
    u_{x}=\frac{u_{1}-u_{2}}{\sqrt{2}},
    \quad
    u_{y}=\frac{u_{1}+u_{2}}{\sqrt{2}},
    \quad
    u_{z}=u_{3}.
\end{equation}
For both plate orientations, the strain tensor in the $\braket{100}$-axes has the form of Eq.~\eqref{eq:strain_decomp}.


\subsection{Dispersive frequency shifts for the coupled Lam\'e and torsional modes}
\label{subsec:explicit}

For the Lam\'e and torsional mode displacement fields, the overlap integrals in Eq.~\eqref{eq:gamma_via_Lambda} can be evaluated analytically. If we disregard the much smaller intrinsic fourth-order elasticity parameters, the scaled frequency shift in Eq.~\eqref{eq:scaled_shifts_basic} for the Lam\'e mode induced by the torsional mode becomes
\begin{subequations}
\label{eq:lame_shift_analytic}
\begin{align}
\aleph_\mathrm{LT}^{\langle100\rangle} &\approx \left(\frac{h_s}{L_s}\right)^2 \frac{6\Lambda_\mathrm{LT}}{C_{11}-C_{12}}, \\
\aleph_\mathrm{LT}^{\langle110\rangle} &\approx \left(\frac{h_s}{L_s}\right)^2 \frac{3\Lambda_\mathrm{LT}}{C_{44}},
\quad \Lambda_\mathrm{LT}\equiv\Lambda^{(4)}_{1166}-\Lambda^{(4)}_{1266},
\end{align}
\end{subequations}
where the superscript is to identify the crystal cut.. The torsional-mode shift induced by the Lam\'e mode is
\begin{subequations}
\label{eq:torsion_shift_analytic}
\begin{align}
\aleph_\mathrm{TL}^{\langle100\rangle} &= \frac{\pi^2\Lambda_\mathrm{LT}}{2C_{44}}, \\
\aleph_\mathrm{TL}^{\langle110\rangle} &= \frac{\pi^2\Lambda_\mathrm{LT}}{C_{11}-C_{12}}.
\end{align}
\end{subequations}
In these equations, $C_{ij} = C\2_{ij} + \Lambda\2_{ij}$ are the components of the full linear elasticity tensor.

Equations~\eqref{eq:lame_shift_analytic} and~\eqref{eq:torsion_shift_analytic} show that the nonlinear frequency shifts of both modes are controlled by the same hole-induced dispersive-coupling coefficient $\Lambda_\mathrm{LT}$. The shift magnitudes differ because this coefficient is divided by different mode stiffness. In the thin-plate limit, the shift of the Lam\'e mode frequency is suppressed by the factor $(h_{s}/L_{s})^2$, which reflects the different geometries of the modes. This gives partial protection of the Lam\'e mode against vibrations of the torsional mode. At the same time, the torsional mode frequency is responsive to vibrations of the Lam\'e mode. Note, however, that we scale the torsional mode amplitude by the plate length, not thickness. 

To realize a highly stable clock, it is important to understand the magnitude and the dependence of the dispersive shift of the both modes on temperature and hole density. They are determined by the coupling parameter $\Lambda_\mathrm{LT}$.

\section{Three-band model and numerical procedure}
\label{sec:three_bands}

To determine the temperature and hole-density dependence of the dispersive mode frequency shifts, we use the same microscopic model, material parameters, and numerical procedure as in Ref.~\cite{Liu2025}, generalized from the single-mode self-Kerr problem to the problem of the Lam\'e--torsional quartic coupling. We treat the valence-band holes within the full three-band Luttinger--Kohn--Bir--Pikus framework appropriate for $p$-doped silicon, where the relatively small spin-orbit splitting makes the split-off band relevant in the temperature and density range of interest. The hole Hamiltonian is written as
\begin{equation}
H_{\mathrm{h}}(\bm{k},\hat{\varepsilon})=
H_{\mathrm{LK}}(\bm{k})+H_{\mathrm{BP}}(\hat{\varepsilon}),
\label{eq:H_LKBP}
\end{equation}
where $H_{\mathrm{LK}}$ describes the heavy-hole, light-hole, and split-off valence bands in the absence of strain, and $H_{\mathrm{BP}}$ is the Bir--Pikus strain Hamiltonian, which is linear in the strain tensor. The explicit expressions for $H_{\mathrm{LK}}$ and $H_{\mathrm{BP}}$ can be found in the literature; see, e.g., Refs.~\cite{Bir1974, Luttinger1955, Winkler2003}. We assume the strain to be a smooth function of coordinates. Therefore, for each wave vector $\bm{k}$, we locally diagonalize the full Hamiltonian, Eq.~\eqref{eq:H_LKBP}, and find three doubly degenerate strain-dependent branches $E_{\nu}(\bm{k},\hat{\varepsilon})$, with $\nu=\{\mathrm{hh},\mathrm{lh},\mathrm{so}\}$ corresponding to heavy-hole, light-hole, and split-off bands, respectively.

The diagonalization gives a strain-dependent hole energy. For the temperatures and hole densities of interest, the dominant contribution to the grand canonical potential $\Omega$ comes from the wave vectors $\bm k$ where the strain-induced correction to the energy is small. Therefore, we expand the energy in a series in $\hat\varepsilon$ as
\begin{equation}
    \label{eq:energy_series}
    E_\nu(\bm k,\hat\varepsilon)= E_\nu^{(0)}+\sum_{n=1}E_\nu^{(n)}(\bm k,\hat\varepsilon),
\end{equation}
where the terms $E_\nu^{(n)}$ are proportional to the convolutions of $n$ tensors $\hat\ep$, that is, to $\|\hat\ep\|^n$.

For each value of the control parameters $T$ and $\nh$, the chemical potential $\mu$ is determined from the condition that the total hole density remains equal to the prescribed value $\nh=-\partial\Omega/\partial\mu$. Therefore $\mu$ is also strain-dependent. Similarly to $E_\nu({\bm k}, \hat\ep)$ this dependence can be described by a series in $\hat{\varepsilon}$,
\begin{align}
\label{eq:chemical_potential}
\mu=\mu^{(0)}+\mathcal{M}^{(1)}_{i}\varepsilon_{i}+\frac{1}{2}\mathcal{M}^{(2)}_{ij}\varepsilon_{i}\varepsilon_{j}+\cdots,
\end{align}
where $\mu^{(0)}$ is the chemical potential in the absence of strain. 

The hole contribution to the elastic part of the free energy density $\Delta\mathcal{F}$ is then obtained order by order by expanding $\Omega$ in Eq.~\eqref{eq:grand_potential} using Eqs.~\eqref{eq:energy_series} and \eqref{eq:chemical_potential}. This gives the tensors $\hat\Lambda\2$ and $\hat\Lambda\4$ in terms of $E_\nu^{(n)}$. The explicit expressions are cumbersome, cf. Ref.~\cite{Liu2025}. The only change relative to the self-Kerr calculation of Ref.~\cite{Liu2025} is that we retain the quartic in $\hat\ep$ terms associated with the combined strain field in Eq.~\eqref{eq:strain_decomp} rather than the pure quartic term of a single mode. In particular, we calculate the quantity that enters the closed-form expressions for the shifts of Sec.~\ref{sec:two_modes}, $\Lambda_\mathrm{LT}=\Lambda^{(4)}_{1166}-\Lambda^{(4)}_{1266}$, which gives the factor multiplying the product of the squared coordinates of the considered modes $Q_{\text{L}}^2 Q_{\text{T}}^2$ in the free energy.

Following the approach of Ref.~\cite{Liu2025}, the integration over $\bm{k}$ in Eq.~\eqref{eq:grand_potential} and in the free-energy expansion is replaced by a summation over a uniform three-dimensional mesh in ${\bm k}$ space for each value of $\hat\ep$, i.e., locally for each ${\bm r}$, which is appropriate given that the wavelengths of the modes of interest are much larger than the hole mean free path. Because the integrands decay exponentially at large $|\bm{k}|$, the integration domain can be truncated; convergence with respect to both the cutoff and the mesh spacing is checked. 

For each mesh point, we diagonalize Eq.~\eqref{eq:H_LKBP} for a discrete set of strain values determined by the mode pattern. To obtain $E_\nu^{(n)}$ in Eq.~\eqref{eq:energy_series}, the resulting band energies are then expanded in the strain using standard finite-difference formulas. These procedures give the same results as a direct perturbation theory, to high accuracy. They allow one to extract the strain expansion of the free energy efficiently without the need of high-order perturbation theory calculations of the eigenvectors.

Once the tensors $\hat\Lambda^{(2)}$ and $\hat\Lambda^{(4)}$ are calculated, the results are combined with the analytic mode profiles of Sec.~\ref{sec:two_modes} to evaluate the change of the mode frequencies $\omega_\alpha$, the Duffing (self-Kerr) nonlinearity coefficients $\gamma^{(\alpha\alpha\alpha\alpha)}$, and the dispersive coupling coefficient $\gamma^{(\alpha\alpha\beta\beta)}$. All silicon band and deformation potential parameters are taken from Ref.~\cite{Winkler2003}, while the temperature dependence of the elastic constants for pure silicon is taken from the data in Ref.~\cite{Varshni1970}. This choice of parameters has been shown to be in excellent agreement with the experimental data on the temperature dependence of the first Lam\'e eigenmode of a silicon plate and with the extension eigenmode of a silicon beam, see Ref.~\cite{Liu2025}. For the results presented below, we use the aspect ratio $h_s/L_s=0.1$, as in the experiment \cite{ComenenciaOrtiz2020}. We also consider the temperature and hole density range where the acceptors are fully ionized, so that $\nh$ is a temperature-independent parameter.


\section{Temperature and density dependence of the dispersive frequency shifts}
\label{sec:results}

\begin{figure}[t]
\centering
\includegraphics[width=8cm]{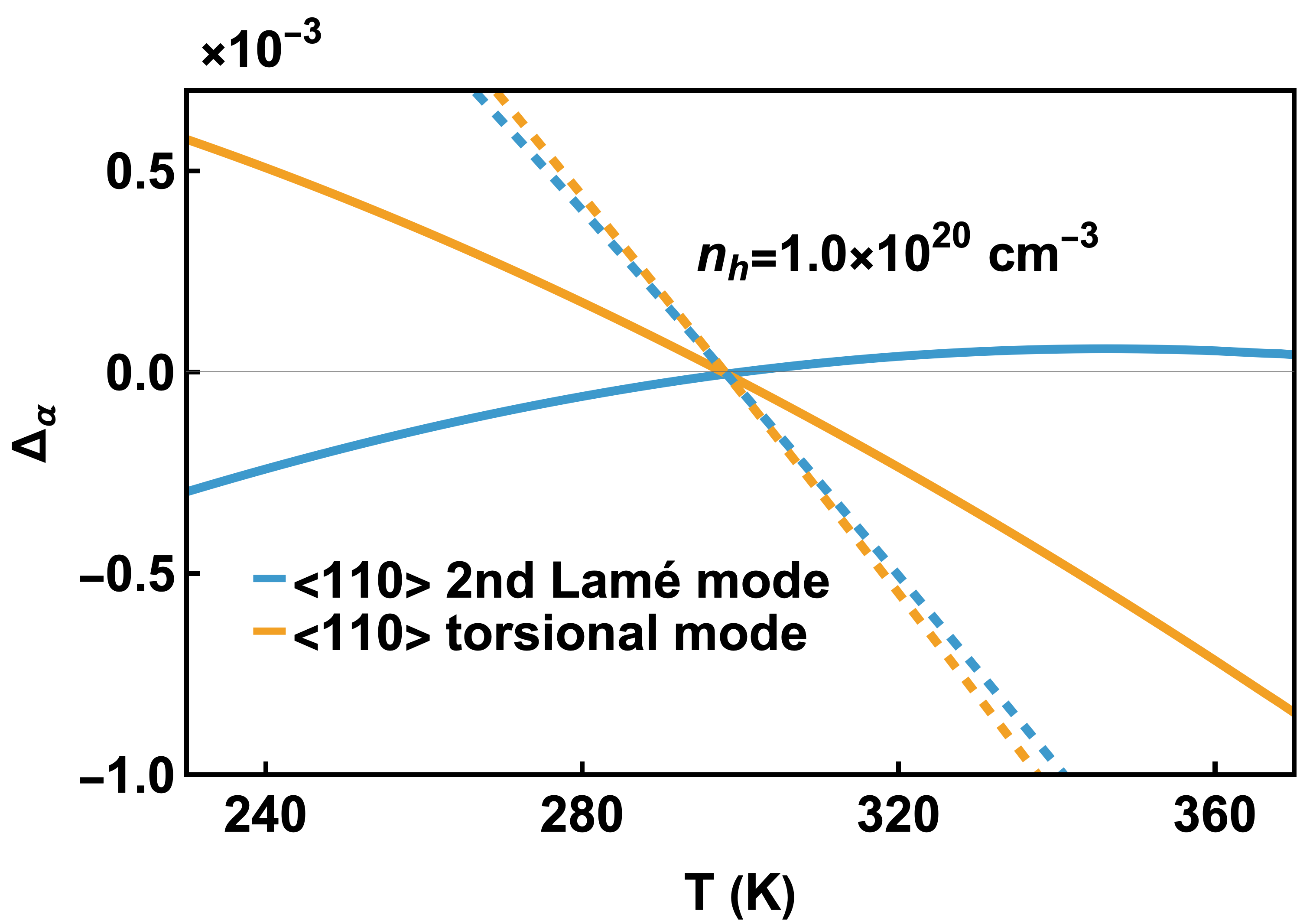}
\caption{Temperature dependence of the eigenfrequencies of the second Lam\'e and torsional modes, $\Delta_\alpha = [\omega_\alpha(T) - \omega_\alpha(T_0)]/\omega_\alpha(T_0)$ for a plate cut out along the $\braket{110}$ axes. The ratio of the thickness to later side of the plate is $h_s/L_s = 0.1$. We set $T_0=298$~K. Solid lines show the temperature dependence in the presence of doping, whereas the results for undoped crystals are shown by dashed lines.}
\label{fig:frequencies}
\end{figure}

We first show in Fig.~\ref{fig:frequencies} the temperature dependence of the frequencies of the second Lam\'e and torsional modes. The results refer to a square $p$-doped silicon plate. The temperature dependence of the eigenfrequencies of the both modes is strongly reduced compared to the undoped case. For the hole density $\nh=10^{20}~\text{cm}^{-3}$, the Lam\'e mode frequency becomes almost fully independent of temperature in a broad temperature range above $300$ K. In contrast, the eigenfrequency of the torsional mode remains strongly temperature-dependent. This underlies the possibility of using this mode as a ``thermometer'' in doped MEMS.

We now present results on the hole-mediated dispersive frequency shifts $\aleph_{\alpha\beta}$, Eqs.~\eqref{eq:lame_shift_analytic} and \eqref{eq:torsion_shift_analytic}, for the Lam\'e and torsional modes of the same system. The main qualitative features of the numerical results can be understood from the analytic expressions~\eqref{eq:lame_shift_analytic} and~\eqref{eq:torsion_shift_analytic}. For the both modes the shifts are controlled by the same hole-induced fourth-order elastic coefficient $\Lambda_\mathrm{LT}$. Therefore the dependencies of $\aleph_\mathrm{TL}$ and $\aleph_\mathrm{LT}$ on temperature are similar. They can be inferred from the estimate \eqref{eq:Lambda_n_est} of $\hat\Lambda\4$. In the nondegenerate regime, i.e., for high temperature, the hole kinetic energy is $E_{\mathrm{kin}}\sim k_B T$, while in the degenerate regime $E_{\mathrm{kin}}=\mu$. Consequently, $\|\hat\Lambda\4\|$, and in particular $\Lambda_\mathrm{LT}$, decrease with the increasing temperature in the nondegenerate regime. However, $\|\hat\Lambda\4\|$ can become nonmonotonic as a function of temperature close to degeneracy, since the chemical potential of a degenerate gas decreases with the increasing temperature. 

If the temperature is fixed, increasing $n_{\mathrm h}$ from small values increases the hole-induced nonlinearity, because more holes are coupled to the modes. However, at sufficiently high densities, the hole gas becomes degenerate and the increasing chemical potential suppresses the nonlinear response. For the quartic-nonlinearity parameters we have
$\|\hat\Lambda\4 \|\propto n_{\mathrm h}/\mu^3\propto n_{\mathrm h}^{-1}$ in the strongly degenerate limit. This leads to a nonmonotonic density dependence of $\Lambda_\mathrm{LT}$ and ultimately of $\aleph_\mathrm{LT}$ and $\aleph_\mathrm{TL}$.


\subsection{Dispersive frequency shifts for fixed hole density}

\begin{figure}[t]
\centering
\includegraphics[width=7cm]{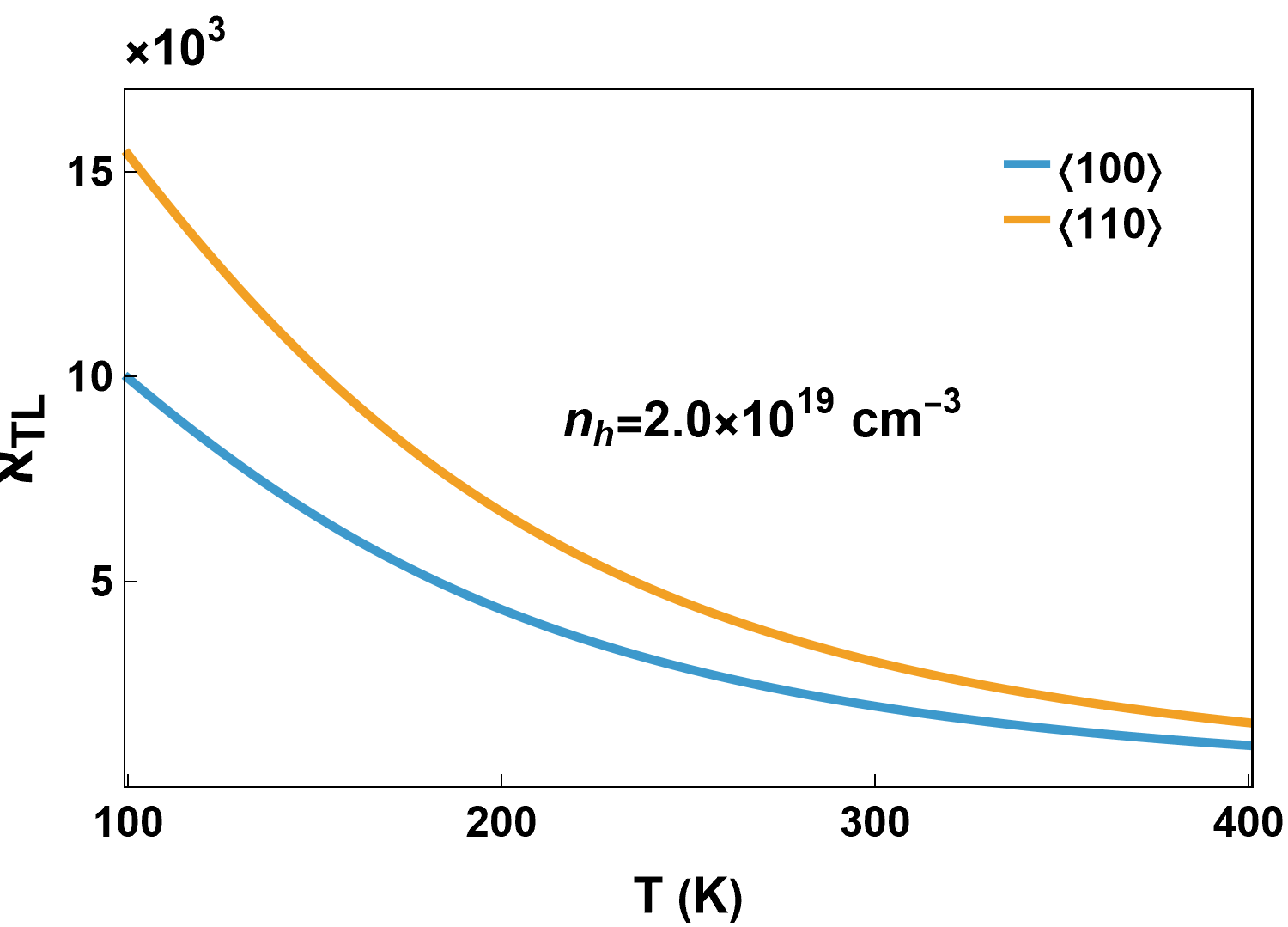}
\vspace{0.5cm}
\includegraphics[width=7cm]{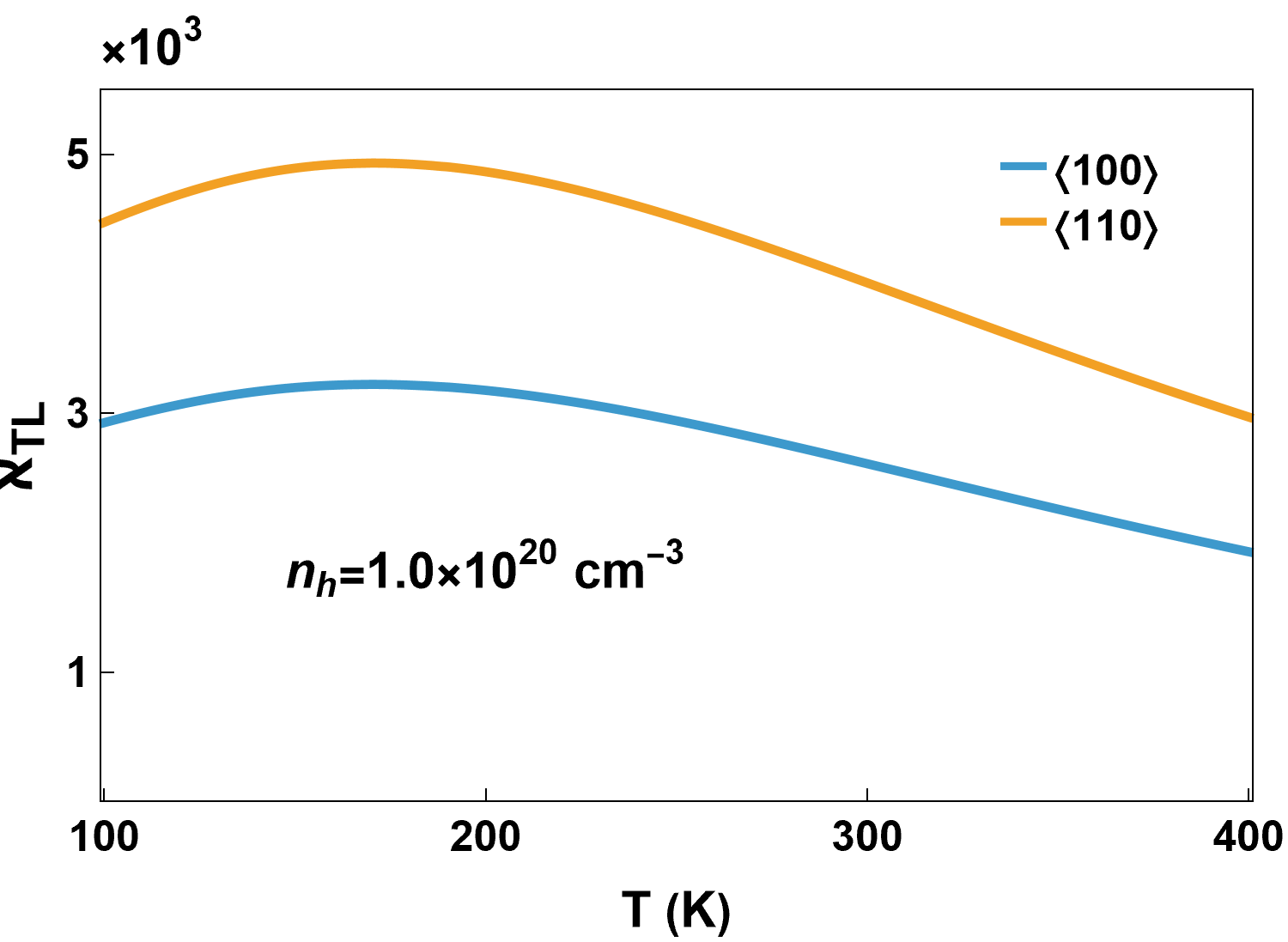}
\caption{Temperature dependence of the dispersive frequency shift $\aleph_\mathrm{TL}$ of the torsional mode scaled by the squared amplitude of the second Lam\'e mode sketched in Fig.~\ref{fig:mode}. The parameter $\aleph_\mathrm{TL}$ is defined in Eq.~\eqref{eq:torsion_shift_analytic}. The results refer to a square single-crystal silicon plate with the ratio of the thickness to length $h_s/L_s=0.1$. The panels show the results for the crystal orientations $\langle100\rangle$ and $\langle110\rangle$. The upper and lower panels refer to the hole density $n_{\text{h}}=2.0\times10^{19}$~cm$^{-3}$ and $n_{\text{h}}=1.0\times10^{20}$~cm$^{-3}$, respectively.}
\label{fig:T_depen_torsional}
\end{figure}

Figure~\ref{fig:T_depen_torsional} shows the temperature dependence of the shift of the torsional mode frequency induced by the Lam\'e mode. At low hole density, $\nh = 2 \times 10^{19}~\mathrm{cm}^{-3}$, the scaled shift decreases monotonically with the increasing temperature, consistent with the estimate $\|\hat\Lambda^{(4)}\|\propto\nh/(k_BT)^3$ for a nondegenerate hole gas. Physically, thermal motion broadens the hole distribution, resulting in a reduced sensitivity to the strain-induced change of the hole energy bands. 

At higher densities, $\nh = 1\times10^{20}~\mathrm{cm}^{-3}$, the temperature dependence becomes nonmonotonic. In this regime, the hole gas is degenerate at low temperature. Since $\mu = E_\mathrm{kin}$ decreases with the increasing temperature, $\Lambda_\mathrm{LT}$ initially increases with the increasing $T$, as seen from Eq.~\eqref{eq:Lambda_n_est}. As the gas becomes nondegenerate for higher $T$, $\Lambda_\mathrm{LT}$ start decreasing. As a result, $\aleph_\mathrm{TL}\propto \Lambda_\mathrm{LT}$ displays a maximum as a function of $T$. The magnitude of $\aleph_\mathrm{TL}$ is large, reaching values $\gtrsim 10^3$ in the considered parameter range. This indicates strong sensitivity of the torsional mode to the Lam\'e mode amplitude.
\begin{figure}[t]
\centering
\includegraphics[width=7cm]{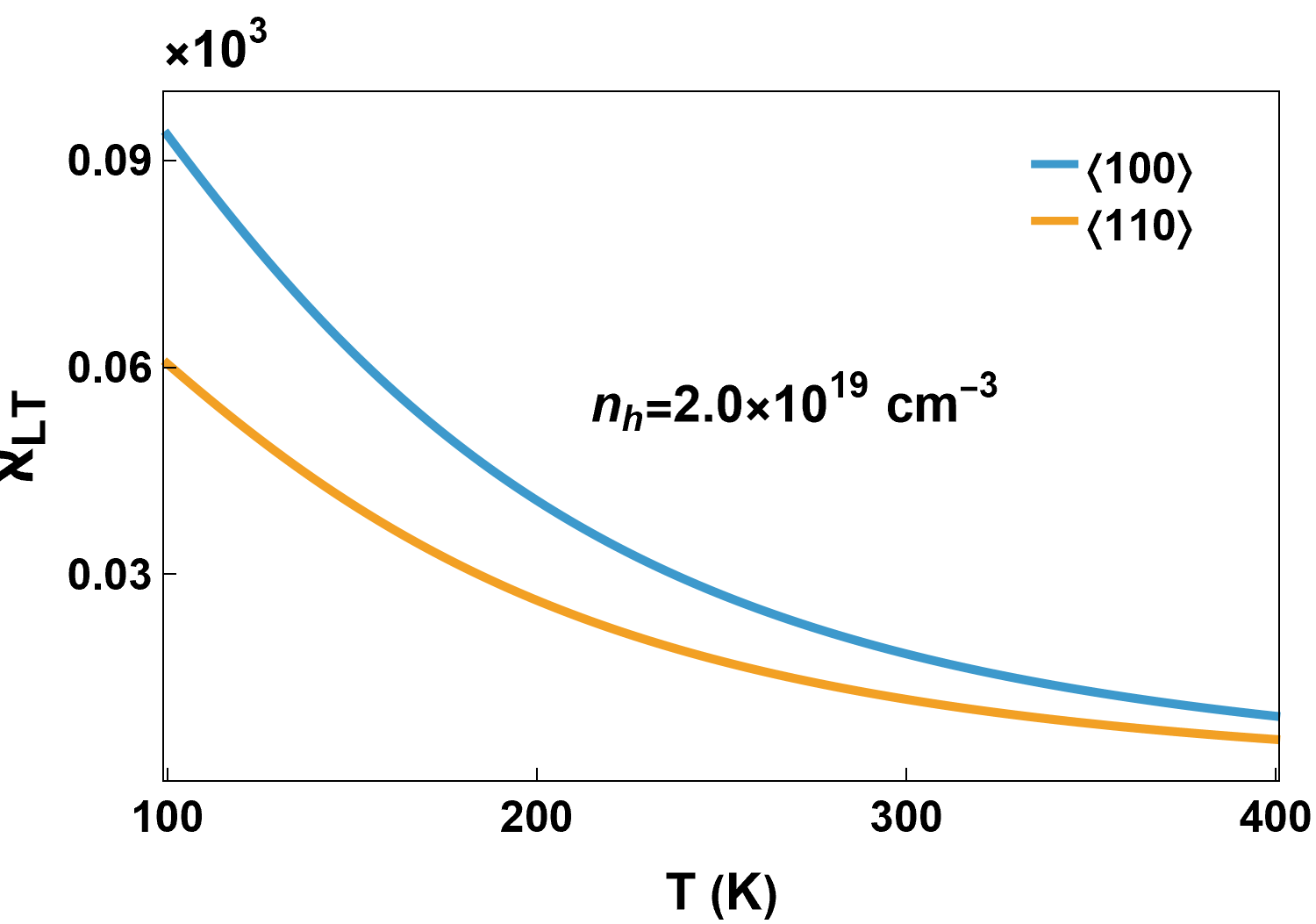}
\vspace{0.5cm}
\includegraphics[width=7cm]{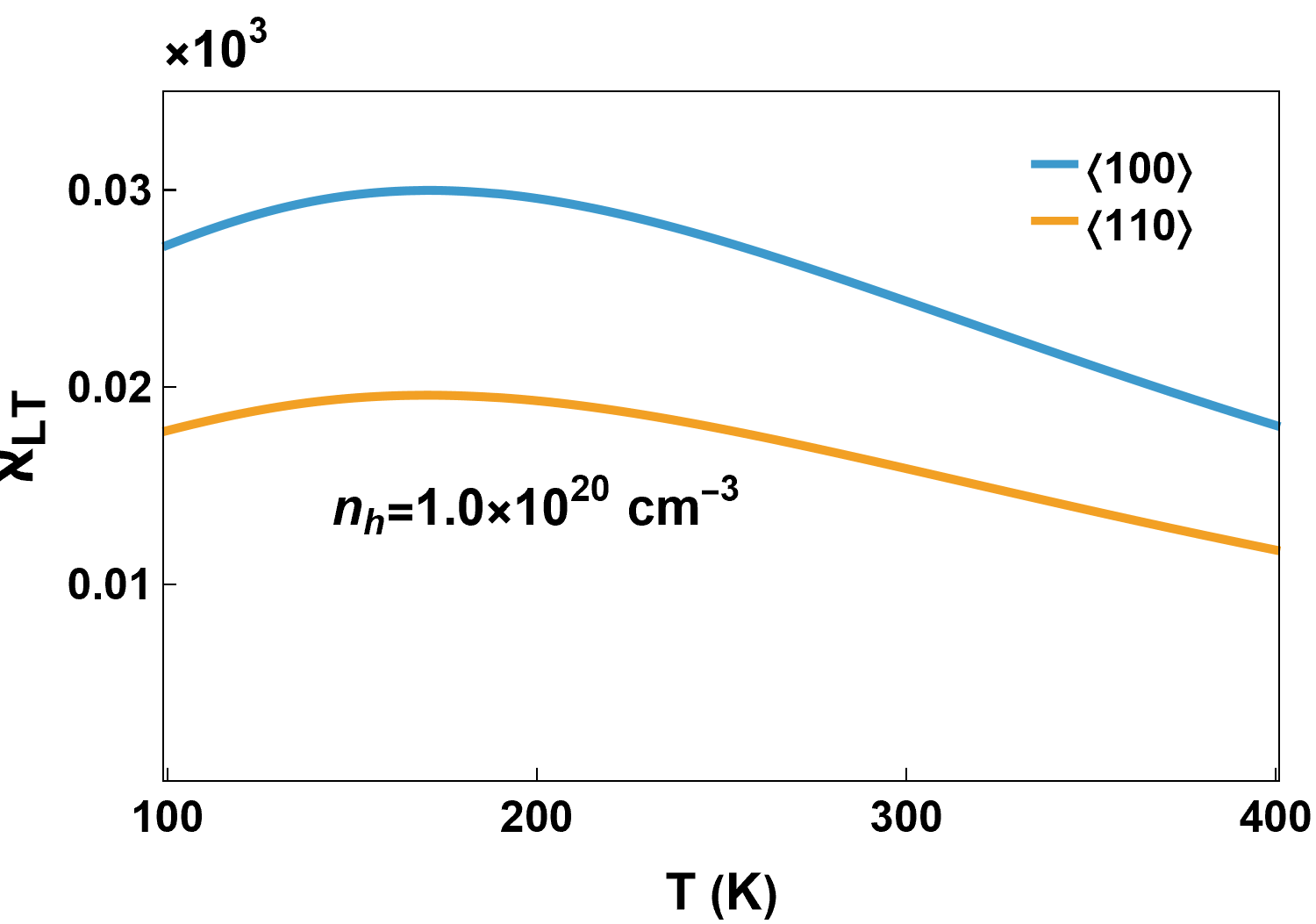}
\caption{Temperature dependence of the dispersive frequency shift $\aleph_\mathrm{LT}$ of the second Lam\'e mode scaled by the squared amplitude of the torsional mode. The modes are sketched in Fig.~\ref{fig:mode}. The parameter $\aleph_\mathrm{LT}$ is defined in Eq.~\eqref{eq:lame_shift_analytic}. The plots refer to a square single-crystal silicon plate with the ratio of the thickness to length $h_s/L_s=0.1$. The panels show the results for the crystal orientations $\langle100\rangle$ and $\langle110\rangle$. The upper and lower panels refer to the hole density $n_{\text{h}}=2.0\times10^{19}$~cm$^{-3}$ and $n_{\text{h}}=1.0\times10^{20}$~cm$^{-3}$, respectively.} 
\label{fig:T_depen_Lame}
\end{figure}

The frequency shift for the studied Lam\'e mode is shown in Fig.~\ref{fig:T_depen_Lame}. Its temperature dependence follows the same trends as in the case of the torsional mode. However, $\aleph_\mathrm{LT}$ is smaller than $\aleph_\mathrm{TL}$ roughly by two orders of magnitude. This has a geometric origin, as seen from Eq.~\eqref{eq:lame_shift_analytic}: the factor $(h_s/L_s)^2$ in this equation is 0.01 for the considered geometry. The torsional mode strain is odd in the direction normal to the plate, and its contribution to the Lam\'e mode frequency shift is small, for a thin-plate limit.

\subsection{Density dependence at fixed temperature}
\begin{figure}[t]
\centering
\includegraphics[width=7cm]{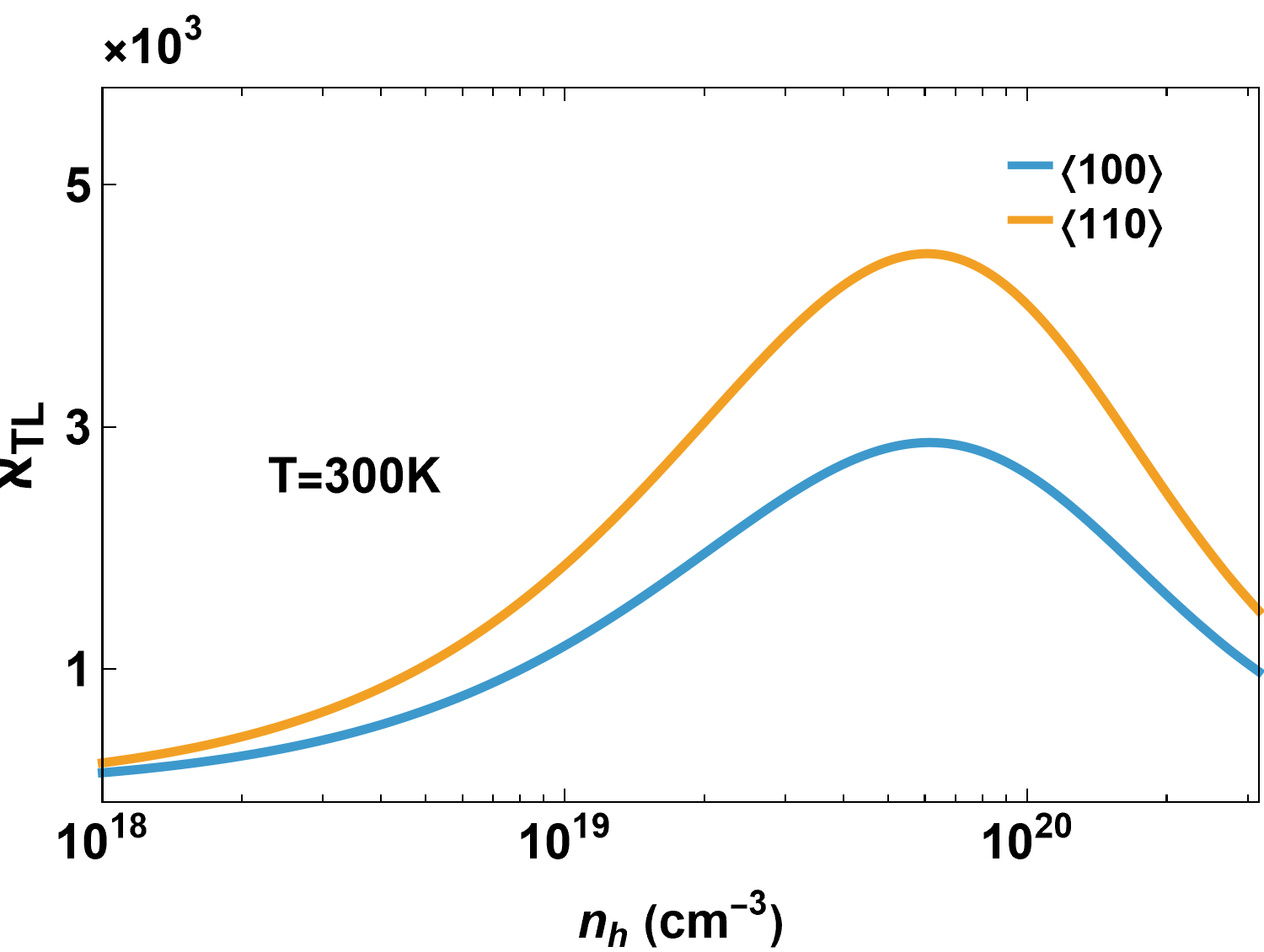}
\vspace{0.5cm}
\includegraphics[width=7cm]{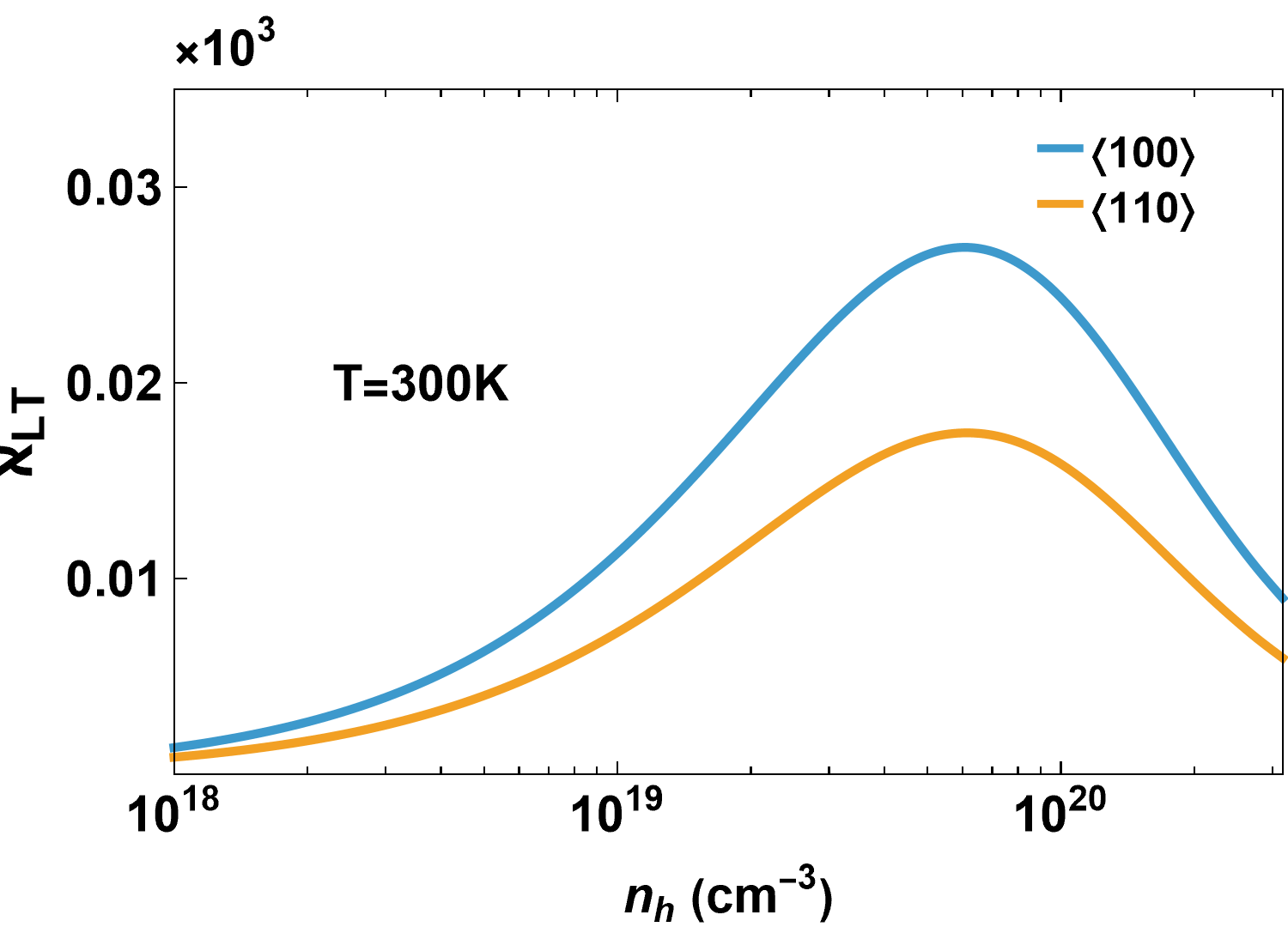}
\caption{Density dependence of the scaled dispersive nonlinear frequency shifts of the torsional (upper panel) and the second Lam\'e (lower panel) modes sketched in n Fig.~\ref{fig:mode}. The shifts are defined in Eqs.~\eqref{eq:torsion_shift_analytic} and \eqref{eq:lame_shift_analytic}, respectively. The plots refer to two crystalline orientations of a square single-crystal silicon plate with the ratio of the thickness to length $h_s/L_s=0.1$. The temperature is $T=300$~K.}
\label{fig:n_depen_two_modes}
\end{figure}

Figure \ref{fig:n_depen_two_modes} shows the density dependence of the dispersive frequency shifts at $T=300~\mathrm{K}$ for the torsional and second Lam\'e modes. Both modes exhibit a strongly nonmonotonic dependence on the hole density $\nh$ for both crystal orientations we studied. The overall behavior is in agreement with what is expected from the qualitative arguments presented above. At low density, where the hole gas is nondegenerate, the shifts grow proportional to the hole density. At large density, they fall off as $\nh^{-1}$, asymptotically. The maxima in Fig.~\ref{fig:n_depen_two_modes} therefore mark the crossover between the nondegenerate and degenerate regimes. We observed the nonmonotonic behavior for all relevant temperatures. The peak moves toward smaller density as temperature decreases, indicating stronger degeneracy-induced suppression of the dispersive frequency shift.


\subsection{Effect of the crystal orientation}
\label{subsubsec:orientation}

Figures~\ref{fig:T_depen_torsional} -- \ref{fig:n_depen_two_modes} show the opposite effect of the nonlinear coupling on the studied two modes: $\aleph_\mathrm{LT}$ is larger for the crystal orientation $\braket{100}$ than for the orientation $\braket{110}$, opposite to the dependence of $\aleph_\mathrm{TL}$ on the crystal orientation. This is a consequence of the difference between the elasticity components $C_{11}-C_{12}$ and $C_{44}$ in Eqs.~\eqref{eq:lame_shift_analytic} and \eqref{eq:torsion_shift_analytic} which, in turn, is due to the different structure of the strain. For a plate whose edges are aligned with the $\langle100\rangle$ axes, in the crystallographic frame, the non-zero strain elements for the Lam\'e mode are $\ep^{(\mathrm{L})}_{xx}$ and $\ep^{(\mathrm{L})}_{yy}=-\ep^{(\mathrm{L})}_{xx}$, whereas the non-zero strain element for the torsional mode is $\ep_{xy}^{(\mathrm{T})}$. In the plate cut along the $\langle110\rangle$ directions, the strain components are effectively interchanged: $\varepsilon_{xx}-\varepsilon_{yy}$ goes over into shear strain component $2\varepsilon_{xy}$ and vice versa.

The orientation dependence has direct implications for dual-mode frequency references. By choosing the plate cut, one can reduce the sensitivity of the frequencies of the torsional and Lam\'e modes to the vibration amplitudes of the Lam\'e and torsional modes, respectively. 


\section{Conclusion}
\label{sec:conclusions}

This paper explores the effect of $p$-type doping of semiconductor micromechanical resonators on the dispersive coupling of their vibrational eigenmodes. We show that doping makes this coupling much stronger than in undoped crystals. This provides an important means of controlling frequencies of one modes by driving other modes. Detailed numerical results refer to a system extensively studied in the experiment, a thin Si plate. Both analytical and numerical results demonstrate that the coupling displays a nonmonotonic dependence on the hole density and, for not too low densities, on the temperature. 

The numerical data refer to the coupling of the second Lam\'e mode and the torsional mode with compatible clamping in a Si plate. This system has a record-high frequency stability \cite{Yan2026}. The primary goal of doping in clock applications is to reduce the temperature dependence of the frequency of the mode that has to be stable and is called a clock mode. Usually this is a Lam\'e mode. The torsional mode is used as a ``thermometer,'' as its frequency is much more sensitive to the temperature variations in the presence of doping. We show that, even though this sensitivity is reduced by doping, it remains sufficiently high in the relevant range of acceptor densities and temperatures. This is in spite of the temperature dependence of the Lam\'e mode frequency being fully compensated in a broad temperature range. 

Besides a comparatively strong coupling of low-frequency eigenmodes of a resonator to each other, doping leads to three-wave and four-wave processes in which there participate both low-frequency eigenmodes and short-wavelength phonons of the crystal, including dispersive coupling to short-wavelength phonons. Such doping-induced processes are effectively limited to the phonons with frequencies small compared to the hole relaxation/thermalization rate. For such phonons, it is stronger than the intrinsic coupling in the host crystal. Still its effect on the eigenfrequencies of low-frequency modes and their temperature dependence is small in a broad temperature range centered at room temperatures. This is because the intrinsic nonlinearity couples low-frequency modes to phonons with thermal wavelengths, which have a much higher density of states than the phonons with frequencies limited by the hole relaxation rate.

An important mesoscopic effect is fluctuations of the eigenfrequencies of low-frequency modes. Because of thermal fluctuations of the amplitudes (occupation numbers) of high-frequency phonons, dispersive coupling to such phonons leads to fluctuations of the frequencies of low-frequency modes. As we show, these fluctuations lead to a very slow phase diffusion in micromechanical systems. However, they may become important for nanoscale systems.

We note in conclusion that, besides changing the elasticity parameters, doping affects the decay rates of low-frequency eigenmodes in micro- and nano-mechanical resonators. A major decay mechanism of such modes in a broad temperature range is the Akhiezer relaxation. Typically, the frequencies $\omega_\alpha$ of the low-frequency modes are small compared to the decay rates of thermal phonons $\Gamma_{\mathrm{ph}}$. Then the Akhiezer relaxation rate is $\propto\omega_{\alpha}/\Gamma_\mathrm{ph}$. This suggests \cite{Liu2025} that, increasing doping, and thus increasing $\Gamma_\mathrm{ph}$, may unexpectedly lower the Akhiezer damping rate. It would be interesting to explore this effect in the experiment.

\section*{Acknowledgments}
We acknowledge partial support from the Defense Advanced Research Projects Agency (DARPA) H6 program under Cooperative Agreement No. HR0011-23-2-0004. We thank Dr. A. A. Shevyrin for helpful comments and suggestions.

\section*{Data Availability}
The source code developed in this study is openly available in the
GitHub repository \cite{Github_Liu_Elewa}.

\end{document}